\documentclass[acmtog]{acmart}
\usepackage{amsmath}
\usepackage{bm}

\DeclareMathOperator*{\argmin}{arg\,min}
\AtBeginDocument{%
  \providecommand\BibTeX{{%
    \normalfont B\kern-0.5em{\scshape i\kern-0.25em b}\kern-0.8em\TeX}}}

\DeclareMathOperator{\grad}{grad}
\DeclareMathOperator{\Hess}{Hess}
\newcommand{\M}{\mathcal{M}}
\newcommand{\TM}{T\!\M}

\newcommand{\originF}{\bar{f}}
\usepackage{cleveref}
\usepackage{algorithm}
\usepackage{algorithmic}
\usepackage{multirow}

\makeatletter
\def\thickhline{%
  \noalign{\ifnum0=`}\fi\hrule \@height \thickarrayrulewidth \futurelet
   \reserved@a\@xthickhline}
\def\@xthickhline{\ifx\reserved@a\thickhline
               \vskip\doublerulesep
               \vskip-\thickarrayrulewidth
             \fi
      \ifnum0=`{\fi}}
\makeatother

\newlength{\thickarrayrulewidth}
\begin{document}

\title{Implicit Bonded Discrete Element Method with Manifold Optimization}

\author{Jia-Ming Lu}
% \affiliation{%

\author{Geng-Chen Cao}
% \affiliation{%

\author{Chen-Feng Li}
% \affiliation{%

\author{Shi-Min Hu}
% \affiliation{%

% \affiliation{%

% \affiliation{%

% \affiliation{%

% \affiliation{%

\renewcommand{\shortauthors}{Jia-Ming Lu, et al.}

\begin{teaserfigure}
    \centering
    \includegraphics[width=0.49\linewidth, trim=600 0 600 800, clip]{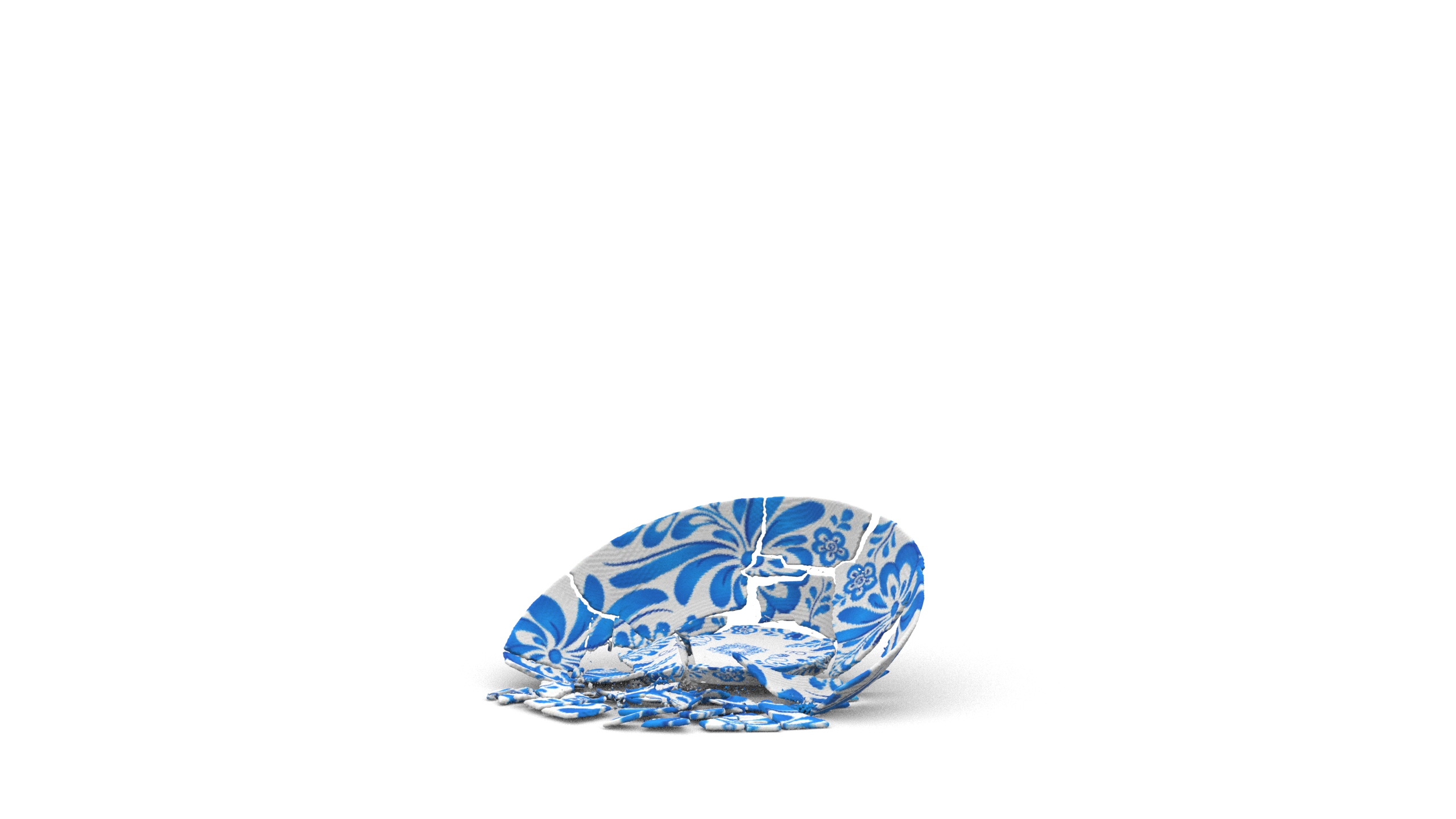}
    \includegraphics[width=0.49\linewidth, trim=400 0 400 200, clip]{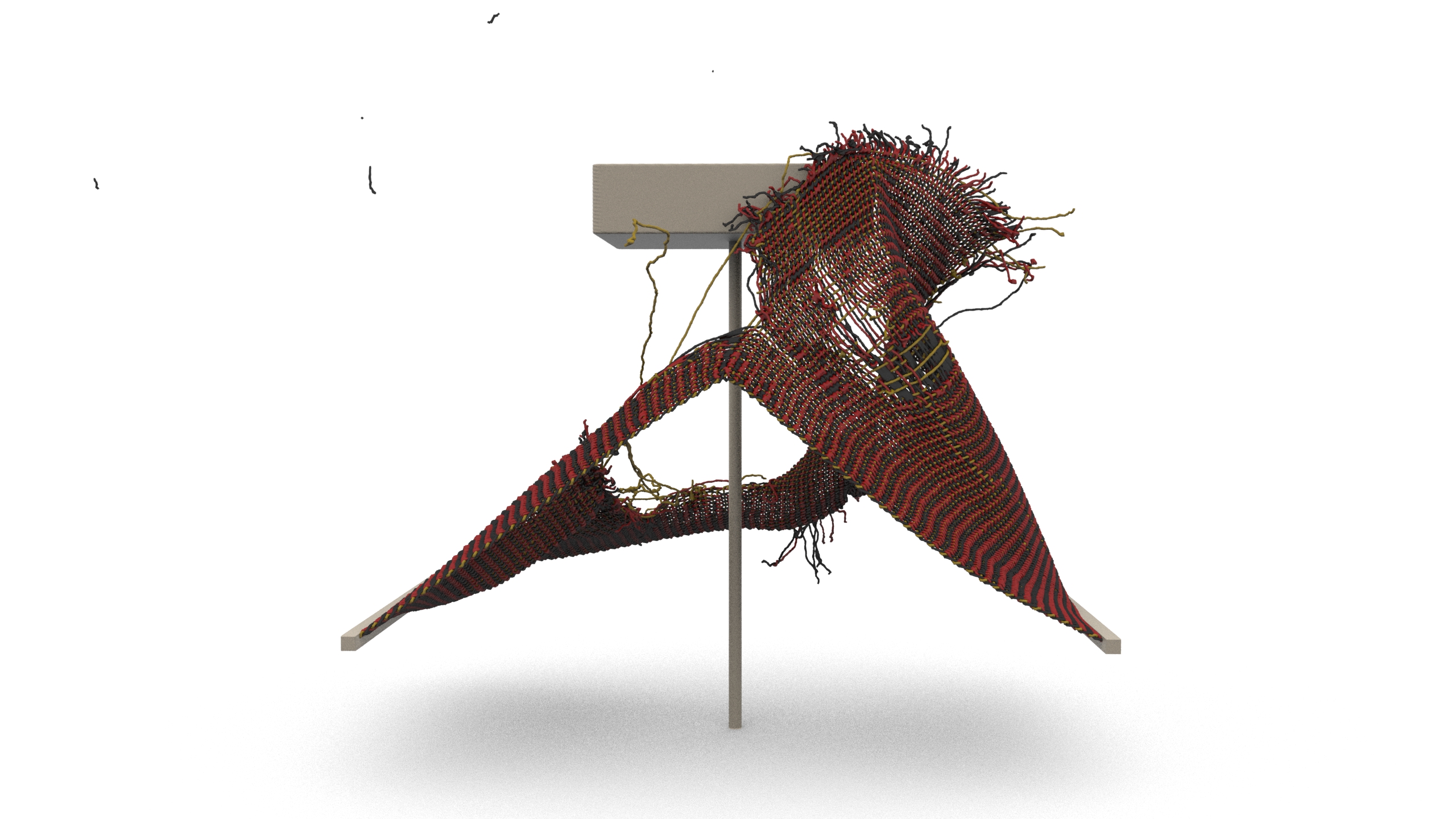}
    \caption{(a) A ceramic plate dropped on a hard floor and broken into pieces. Our simulation captures a great level of details using only 219K particles, with a 2.4x speedup over the state-of-the-art method. (b) A piece of knitted fabric being torn apart, demonstrating our algorithm's ability to handle extreme deformations and material failure in textiles, with a 3.3x speedup over the state-of-the-art method.}
    \label{fig:plate_cloth}
\end{teaserfigure}
\begin{abstract}
This paper proposes a novel simulation approach that combines implicit integration with the Bonded Discrete Element Method (BDEM) to achieve faster, more stable and more accurate fracture simulation. The new method leverages the efficiency of implicit schemes in dynamic simulation and the versatility of BDEM in fracture modelling. Specifically, an optimization-based integrator for BDEM is introduced and combined with a manifold optimization approach to accelerate the solution process of the quaternion-constrained system. Our comparative experiments indicate that our method offers better scale consistency and more realistic collision effects than FEM and MPM fragmentation approaches. Additionally, our method achieves a computational speedup of $2.1 \sim 9.8$ times over explicit BDEM methods.
\end{abstract}

\begin{CCSXML}
<ccs2012>
<concept>
<concept_id>10010147.10010371.10010352.10010379</concept_id>
<concept_desc>Computing methodologies~Physical simulation</concept_desc>
<concept_significance>500</concept_significance>
</concept>
</ccs2012>
\end{CCSXML}

\ccsdesc[500]{Computing methodologies~Physical simulation}

\keywords{Discrete Element Method, Fracture Simulation, Optimization-based Integrator, Manifold Optimization}

\maketitle

\section{Introduction}
\label{sec:Introduction}

Fracture and damage occur to all materials, and they represent a fundamental type of interaction in virtual environments. As an important topic in physics-based simulation, fracture modelling is a challenging task due to the complex underlying physical processes. The ubiquity of fractures in real life sets a high standard for plausible results to satisfy human perception. To date, most simulation methods for materials are based on the continuum theory, which have limited capacity in capturing fracture phenomena. The BDEM can better simulate fractures \cite{lu2021simulating}. One of the most significant challenges in fracture simulation is keeping the error of internal forces under control whilst allowing an acceptable error rate in fracture propagation. For simulation methods based on the continuum theory, the accuracy in stresses is typically much poorer than the accuracy in displacements, making it hard if not impossible to manage the simulation error during fracture propagation. The BDEM does not have this limitation, as its stress and displacement are not hardwired through differential operators.

However, solving BDEM systems is complicated by the stiff nature of the unit length constraint in quaternion systems, and it becomes more challenging for such materials as rope and cloth due to the added complexity of shear stiffness. The collision stiffness in discrete element simulations is another major challenge, as it can significantly increase the number of Newton iterations required. These issues are prevalent and well recognized in BDEM simulations \cite{andre2012discrete, nguyen2021discrete}. Implicit integrators are widely used in computer graphics to accelerate the efficiency of dynamic simulations, where the backward Euler scheme is the most prevalent method \cite{baraff1998large, hirota2001implicit, volino2001comparing, martin2011example, liu2013fast}. Other integrators, such as the implicit-explicit method \cite{eberhardt2000implicit, stern2009implicit} and exponential integrators \cite{michels2014exponential, chen2017exponential}, have also received significant attention. Recent work used a variational integrator to model the inter-granule contact \cite{de2022variational}. However, none of these well-established implicit schemes is effective for BDEM. 

To address this challenge, we introduce an optimization-based integrator to reformulate BDEM and combine it with a manifold optimization approach to deliver faster, more stable and more accurate fracture simulations. Our approach leverages the strength of BDEM in fracture modelling, whilst achieving superior computational efficiency with the resulting implicit solution scheme. Our experiments show that without loss of accuracy, the proposed approach is more efficient than explicit BDEM methods. The main contributions include:
\begin{itemize}
    \item
    A stable optimization-based integrator suitable for implicit BDEM is formulated, and it allows large time steps. 
    \item
    A manifold optimization approach for BDEM is introduced to speed up the solution of quaternion-constrained systems. 
\end{itemize}

The rest of the paper is organized as follows. \S \ref{sec:Related Works} briefly recaps related works in fracture simulation and discrete element methods. The main body of work is presented in \S \ref{sec:Implicit Bonded Discrete Element Method}, \S \ref{sec:Manifold Optimization}, which present respectively the optimization-based integration for BDEM and the manifold optimization approach to accelerate the solution process. More implementation details are provided in \S \ref{sec:Implementation Details}, followed by results and discussions in \S \ref{sec:Results and Discussion}. Finally, concluding remarks are made in \S \ref{sec:Conclusion}. 

\section{Related Works}
\label{sec:Related Works}

\paragraph{Fracture Simulation}
Fracture simulation is an important topic in computer graphics, owing to the ubiquity of fracture phenomena in real life. A number of non-physical methods have been proposed to produce crack patterns \cite{raghavachary2002fracture, hellrung2009geometric, su2009energy, bao2007fracturing, zheng2010rigid, schvartzman2014fracture, muller2013real}, whilst physics-based methods are more accurate and have become the preferred approach for fracture simulation. Till recently, the Finite Element Method (FEM) is the most widely used physics-based technique for fracture simulation \cite{o1999graphical, muller2001real, wicke2010dynamic, koschier2014adaptive, pfaff2014adaptive}. Other physics-based methods include the Extended Finite Element Method (XFEM) \cite{kaufmann2009enrichment,chitalu2020displacement}, the Boundary Element Method (BEM) \cite{hahn2015high, hahn2016fast, zhu2015simulating}, the mesh-less method \cite{pauly2005meshless}, peridynamics \cite{chen2018peridynamics}, and the Material Point Method (MPM) \cite{wolper2019cd, wolper2020anisompm,fan2022simulating}. Each of these methods has its own pros and cons and is suitable for different fracture scenarios. Besides addressing underlying physics, accurately capturing fracture surfaces is crucial for realistic visual effects in fracture propagation. For FEM fracture simulation, methods like adaptive remeshing \cite{chen2014physics,koschier2014adaptive,pfaff2014adaptive,wicke2010dynamic}, duplicate elements \cite{molino2004virtual}, and sub-elements \cite{nesme2009preserving,wojtan2008fast} are employed to capture and prevent poor-quality elements. Explicit surface methods \cite{koschier2017robust}, post-processing techniques \cite{wang2019simulation,chitalu2020displacement,kaufmann2009enrichment}, level-set methods \cite{hahn2015high}, and mesh-based methods \cite{chitalu2020displacement,sifakis2007arbitrary,zhu2015simulating} offer alternative approaches for accurately representing fracture surfaces.

\paragraph{Discrete Element Methods} 
Introduced by Cundall and Strack for solving problems in rock mechanics \cite{cundall1971,cundall1979}, the Discrete Element Method (DEM) is widely used in engineering for modelling granular materials. The DEM has also been adopted in computer graphics for visual effects involving granules and particles \cite{bell2005particle,alduan2009simulation,rungjiratananon2008real,yue2018hybrid}. To simulate continuum media, Potyondy and Cundall extended the DEM to form the BDEM \cite{potyondy2004bonded}, and since then it has seen growing adoption in various engineering applications. Examples include \cite{andre2012discrete}, where a cohesive bond model was used to represent continuum objects, and \cite{nguyen2021discrete}, where a plastic cohesive bond was used to simulate ductile fractures. The BDEM was recently introduced by \cite{lu2021simulating} to the computer graphics community, and it was proven to be robust and flexible for producing realistic visual effects involving fractures. As far as plausible visual effects are concerned, the main limitation of BDEM fracture simulation is its computational efficiency, due to the need for small time steps in the simulation. 

\paragraph{Simulation with Rotation}
Different from FEM, where the degrees of freedom of motion are defined on a set of points / nodes and include only translations, the degrees of freedom for DEM are defined on rigid spheres, which include both translational and rotational degrees of freedom. The explicit inclusion of rotation as part of element states is a major advantage for DEMs to represent the deformation and motion of real world objects. There are other methods that also explicitly account for the rotational motion. A prominent and well-established category within this domain is the simulation of rigid body motion, which has been subjected to extensive exploration for several decades \cite{stewart2000rigid, baraff1989analytical,bender2014interactive,hahn1988realistic}. Recent endeavors in rigid body simulation include the employment of the Extended Position-Based Dynamics method for solving the motion of rigid bodies \cite{muller2020detailed}, differentiable rigid contact models \cite{geilinger2020add}, and the modeling of rigid body motion and collision resolution through the formulation of incremental potential \cite{ferguson2021intersection}. Certainly, in addition to rigid body simulations, the incorporation of rotation as a degree of freedom is also utilized in various other methods. One such method uses particles with orientation \cite{muller2011solid}, but this geometric approach does not capture basic physical concepts such as momentum and stress, which can limit its ability in producing realistic visual effects. Another example is rod elements for modelling slender objects. The Discrete Elastic Rod (DER) approach \cite{bergou2008discrete} was used to compute rod curvature, whilst Cosserat rods were used in \cite{kugelstadt2016position,soler2018cosserat} to model complex bending and torsion effects. The rod elements approach is effective for slender and flexible structures (e.g. ropes and fibers), but they are less effective for simulating generic objects. 

\paragraph{Optimization-based Integrator}

In physics-based simulations, the resolution of complex nonlinear equations is often required, which can lead to inefficient and unstable solutions due to their inherent complexity. An integrator based on energy optimization is a method designed to enhance the efficiency and stability of solving nonlinear equations. This approach has been applied to various types of computational problems, including fluid dynamics \cite{weiler2016projective}, crowd simulation \cite{karamouzas2017implicit}, and elastic bodies \cite{gast2015optimization,martin2011example,fratarcangeli2016vivace,wang2016descent} and nonlinear materials \cite{li2019decomposed, loschner2020higher, kee2023optimization}. Projective dynamics represent a class of optimization-based methods for solving constrained systems, capable of rapidly resolving a diverse array of materials \cite{bouaziz2014projective,liu2013fast,liu2016towards,narain2016admm,weiler2016projective}. 
Incremental potential is also a class of methods based on energy optimization, which has recently been employed to handle collisions between various objects \cite{li2020incremental,lan2021medial,li2020codimensional,ferguson2021intersection}. By introducing the interior point method to resolve impacts, it consistently achieves interpenetration-free outcomes. In our work, akin to the approach presented in \cite{ferguson2021intersection}, we have derived an integrator with rotational degrees of freedom in an energy-optimized form. In contrast to the complex formulation originally targeted at rigid body simulation, our integrator has been specifically tailored for the BDEM system, resulting in a significantly streamlined solution that achieves efficient computation.

\begin{figure*}[ht]
    \centering
    \includegraphics[width=0.33\linewidth, trim=350 050 350 350, clip]{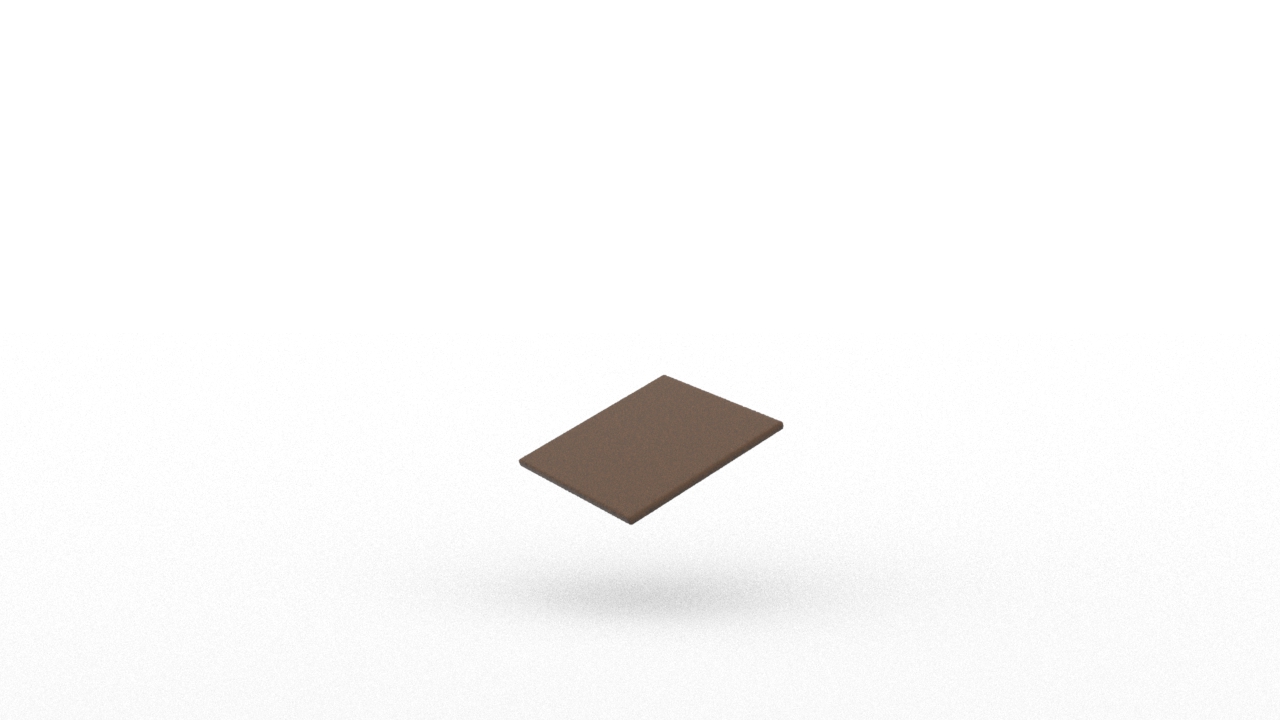}
    \includegraphics[width=0.33\linewidth, trim=350 050 350 350, clip]{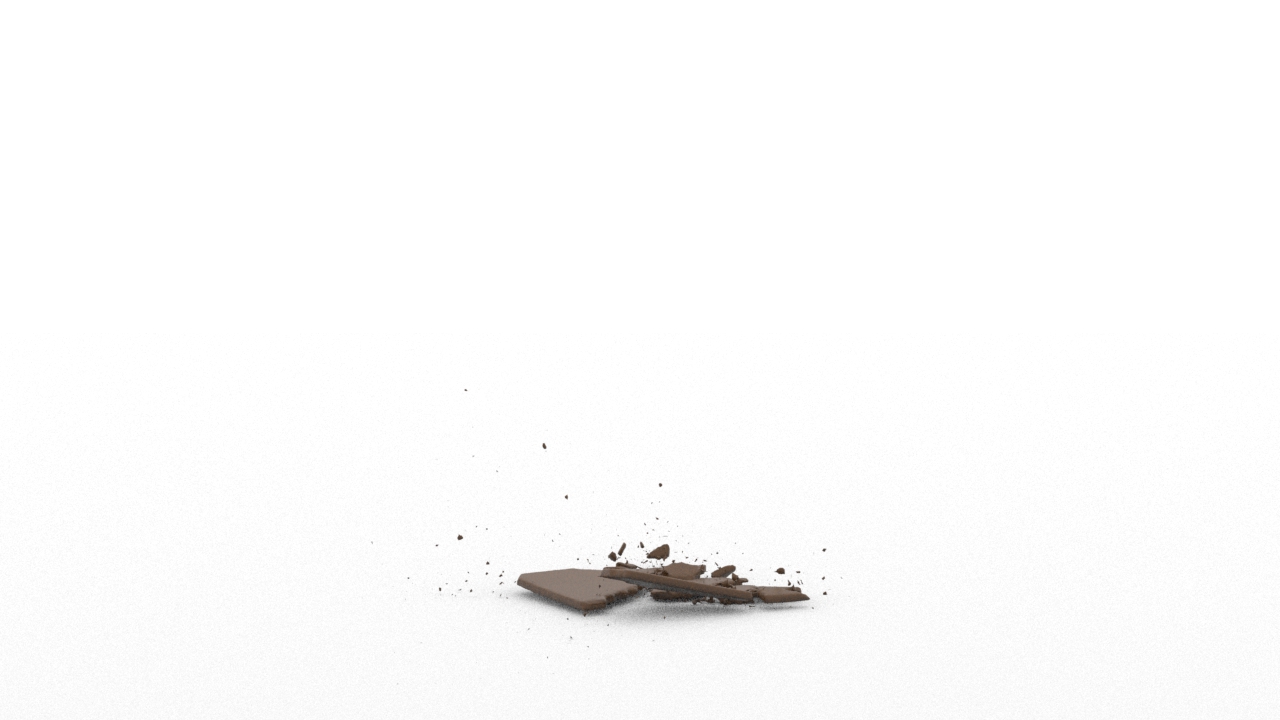}
    \includegraphics[width=0.33\linewidth, trim=350 050 350 350, clip]{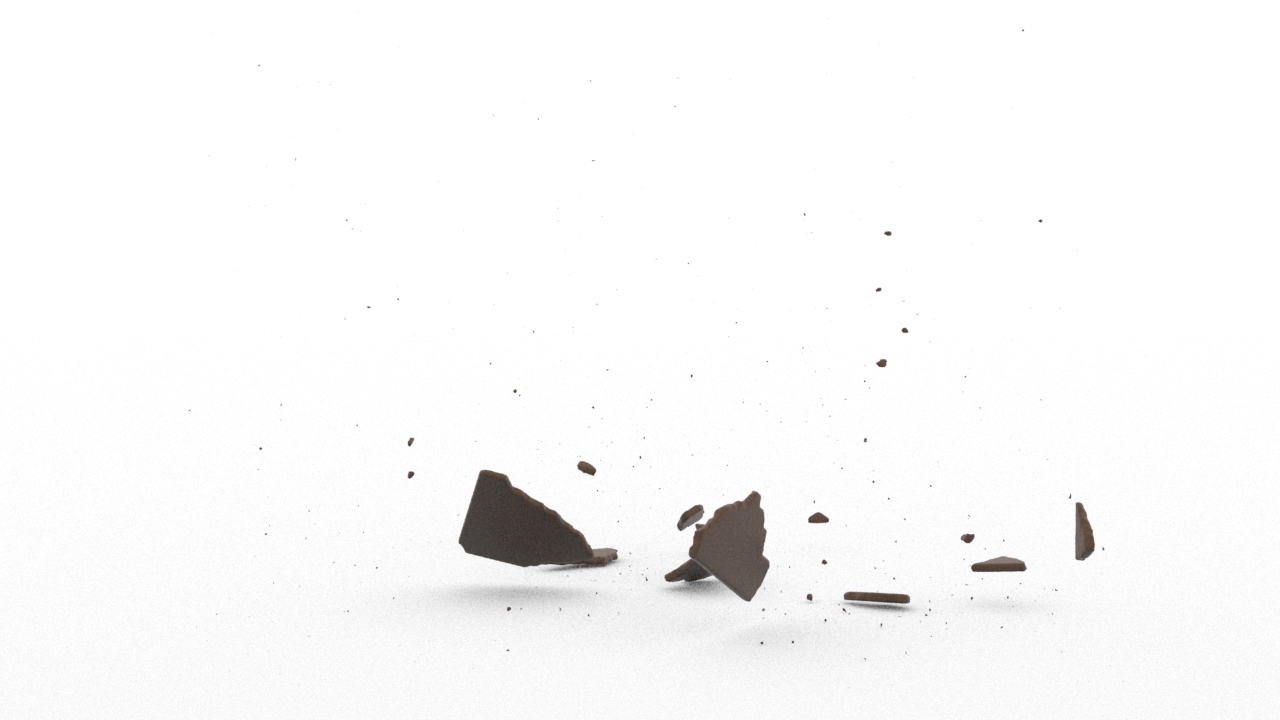}
    \caption{showcases the experimental results of our method on a chocolate drop impact simulation. The figure illustrates the moment when a chocolate drop hits the ground and cracks into several pieces and small fragments. The simulation shows the effectiveness of our approach in accurately modeling high-speed object collisions and fracture phenomena.}
    \label{fig:chocolate}
\end{figure*}

\section{Implicit Bonded Discrete Element Method}
\label{sec:Implicit Bonded Discrete Element Method}

A significant difference between BDEM and other approaches is the direct consideration of rotational Degrees of Freedom (DoFs) as system variables. Previous methods solve BDEM systems using explicit formulations that store rotation status as quaternions, Euler angles or axis angles, and compute torques and angular velocities explicitly to alter the rotation status of discrete elements. In implicit formulations, position and rotation states cannot be explicitly updated, and forces or torques are often implicit or invisible during the solution process. To form an implicit BDEM scheme, it requires a different formulation.

Solving generic nonlinear equations are unstable, a promising strategy to tackle this challenge is transforming the nonlinear simulation problem into a nonlinear optimization problem that is easier to solve \cite{gast2015optimization}. Optimization-based integrators are effective in improving the stability and convergence speed of nonlinear solution processes. Considering BDEM system, a direct idea is that we formulate the solution of motion equations with rotational degrees of freedom into an energy minimization form to achieve a fast and stable integrator. In conventional rigid body solutions, due to the introduction of rotational degrees of freedom, the mass matrix $M$ in the extended motion equations is varies with the motion state \cite{liu2012quick,geilinger2020add}. Therefore, we cannot directly write the extended motion equations in the form of energy minimization. In pursuit of an improved solution form, akin to the approach detailed in \cite{kane2000variational}, we have re-derived the motion equation for BDEM based on quaternions. We have discovered that, leveraging the inherent characteristics of BDEM, we can ultimately arrive at an exceedingly succinct formulation for optimization-based integrator.

\subsection{Optimization-based Integrator}
Building upon Hamiltonian mechanics for quaternion-based rigid body dynamics, we present a simplified computational framework for BDEM simulations of spherical particles. Our key contribution is the derivation of a constant mass matrix formulation that maintains solution convergence while eliminating the need for rotation state-dependent mass matrices. The complete mathematical foundation is presented in ~\Cref{sec:discrete_lagrangian}. Our formulation yields a block-diagonal extended mass matrix $M_s$:
\begin{equation}
    M_s=\begin{pmatrix}
        m I_3 &\\
        & \frac{8}{5}mr^2 I_4
    \end{pmatrix}.
\end{equation}
Where $m$ is the particle mass, $r$ is the sphere radius, and $I_3$ and $I_4$ are identity matrices of dimensions 3 and 4, respectively.

This constant mass matrix enables us to express the dynamics as a discrete variational principle, leading to the following optimization formulation:
\begin{equation}
\Phi(\bm x_{i+1}) = \frac{1}{2\Delta t^2}(x_{i+1}-\hat{z})^T M_s (x_{i+1}-\hat{z})+V(\frac{x_{i+1}+x_i}{2}).
\end{equation}
where $\Delta t$ is the time step size, $x_i$ represents the system state at time step $i$ (including both position and quaternion components), and $V$ denotes the potential energy. The auxiliary variables $\hat{x}$ and $\hat{z}$ are defined as:
\begin{equation}
\begin{split}
\hat{x} &= 2x_i-x_{i-1} \\
\hat{z} &= \hat x - \frac{1}{2}M_s^{-1}\Delta t^2\nabla V(\frac{x_{i-1}+x_i}{2})
\end{split}
\end{equation}
These variables are computed from known states at previous time steps. The system evolution is then determined by solving the constrained optimization problem:
\begin{equation}
x_{i+1} = \argmin_x \Phi(x) \quad \text{s.t.} \quad C(x)=0,
\end{equation}
where $C(x)$ represents the quaternion unit length constraint (see \Cref{appendix:nullspace_operator}).
This variational integrator, a symplectic integrator with second-order accuracy, is now ready for our simulation and compatible with our following manifold optimization approach. However, to further reduce computational costs, we introduce an approximation in our experiments to tradeoff the computational speed and simulation accuracy.
We approximate both $\nabla V(\frac{x_{i+1}+x_i}{2})$ and $\nabla V(\frac{x_{i-1}+x_i}{2})$ with $\nabla V(x_{i+1})$. This approximation, while reducing the integrator to first-order accuracy, maintains compatibility with our manifold optimization approach. Empirical comparisons reveal that the primary drawback is slightly increased damping, a trade-off for improved computational efficiency. The modified form, similar to the implicit Euler method, is:
\begin{equation}
\label{eqn:implicit_euler_integrator}
\frac{1}{\Delta t^2}M_s(x_{i+1}-\hat x)+\nabla V(x_{i+1})=0.
\end{equation}
with the corresponding objective function:
\begin{equation}
\label{eq:optimization form}
\Phi(\bm x_{i+1}) = \frac{1}{2\Delta t^2}(\bm x_{i+1}-\hat{\bm x})^T M_s (\bm x_{i+1}-\hat{\bm x})+V(\bm x_{i+1}).
\end{equation}
Note that this formulation bears a striking resemblance to its counterpart without rotational degrees of freedom. This similarity might raise questions about the treatment of rotation-related terms. To address this potential source of confusion and provide a comprehensive understanding, we present an alternative derivation with a detailed exposition of the rotational dynamics in \Cref{sec:Hamilton_System}. 

\begin{figure*}[ht]
    \centering
    \includegraphics[width=0.33\linewidth, trim=450 0 450 450, clip]{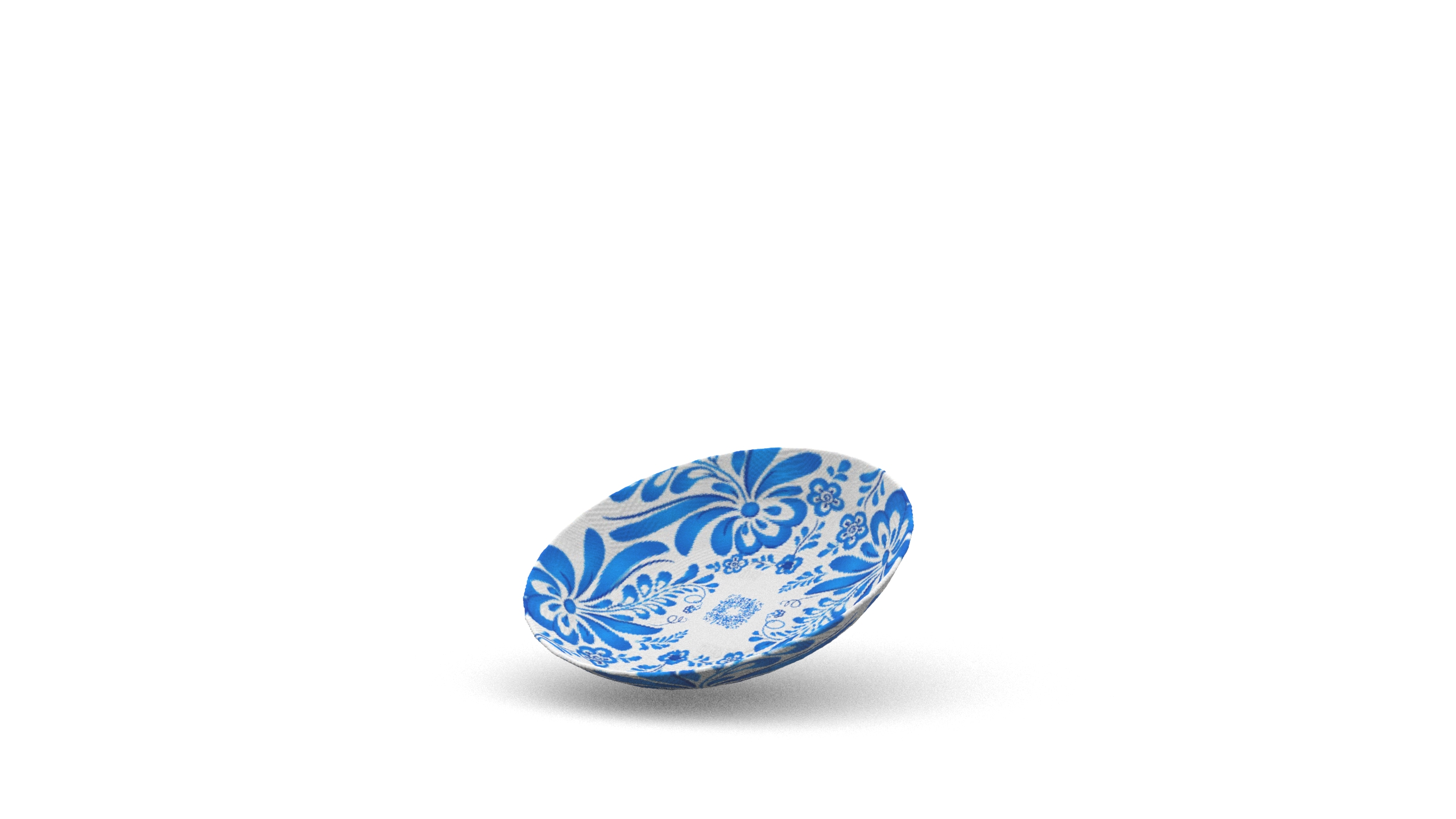}
    \includegraphics[width=0.33\linewidth, trim=450 0 450 450, clip]{figures/plate_middle.jpg}
    \includegraphics[width=0.33\linewidth, trim=450 0 450 450, clip]{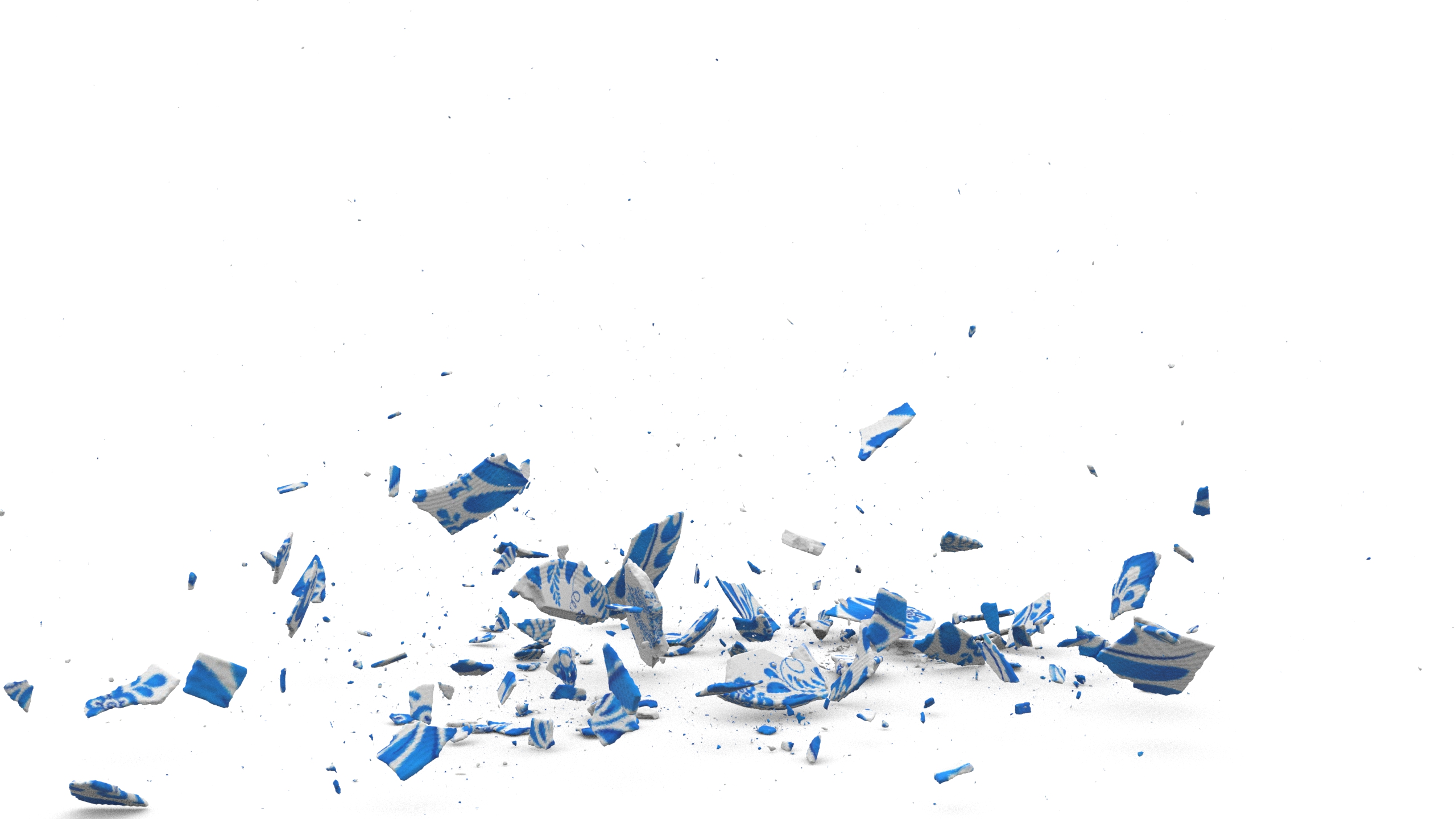}
    \caption{A ceramic plate dropped on a hard floor and broken into pieces. Our simulation captures highly detailed fractures using only 219K particles, with a 2.4x speedup over the state-of-the-art method.}
    \label{fig:plate}
\end{figure*}

\subsection{Potential Energy in BDEM}
Now we require only the specification of a potential function to solve the BDEM system. In a setup where two vertices $x_i$=$\{p_i,q_i\}$ and $x_j$=$\{p_j,q_j\}$ are bonded together, the deformation of the bond can be decomposed into four forms: stretch, shear, bend, and twist, akin to the form presented in \cite{lu2021simulating}. It is straightforward to derive the bond energy with these four deformation types: 
\begin{equation}
    V = V_{\mbox{stretch}} + V_{\mbox{shear}} + V_{\mbox{bendtwist}}.
\end{equation}

\paragraph{Stretch}
The stretch energy of a bond can be modeled using a simple spring model and expressed as follows:
\begin{equation}
    V_{\mbox{stretch}} = \frac{1}{2}k_n (\|p_i - p_j\| - l_0)^2,
\end{equation}
where $k_n = \frac{E S}{l_0}$, $l_0$ is the original bond length, $E$ is Young's modulus, and $S$ is the bond's cross-sectional area, with $S = \pi (\frac{r_i + r_j}{2})^2$, using the radii $r_i$ and $r_j$ of two bonded elements.

\paragraph{Shear}
The rotation of two elements can be decomposed into a common rotational component and a differential rotational component. As elucidated in the article \cite{lu2021simulating}, the shear energy is modeled by the variation of the common rotational component relative to the rest state of the bond. To extract the common rotation, the equation $q_c = (q_i + q_j)/\|q_i + q_j\|$ is used, where $d = q_c \odot d_0$ represents the current direction. The corresponding energy can be expressed as
\begin{equation}
    V_{\mbox{shear}}=\frac{1}{2}k_t \theta^2,
\end{equation}
where $\theta=\left <d_0,d_q\right >$ denotes the angle between the bond's initial direction $d_0$ and current direction $d$. The shear stiffness is $k_t = G S / l_0$, where $G$ represents the material's shear modulus.

\paragraph{Bend and Twist}
The bend and twist energy of discrete elements can be defined as the difference in their rotational component. It is important to note that the stiffness of bend and twist have distinct relationships with material properties, thus requiring separate decomposition of the two types of rotations. The difference of quaternion, expressed in the form of $G_l(q_j) q_i$, can be transformed to the local frame via the rotation matrix $R_0 = G_l(q_0) G_r(q_0)$, where $q_0$ is the rotation between $d_0$ and $\{1, 0, 0\}$. The resulting vector can then be represented as $t = R_0 G_l(q_j) q_i$. With the aid of the stretch and shear stiffness matrix denoted as $K$, the energy form can be expressed using
\begin{equation}
    \label{eqn:bendtwist_potential}
    V_{\mbox{bendtwist}} = \frac{1}{2} t^T K t.
\end{equation}
The stiffness matrix is defined by
\begin{equation}
    K = \frac{1}{l_0}\begin{pmatrix}
        GJ & & \\
        & EI & \\
        & & EI 
    \end{pmatrix},
\end{equation}
here, $I=\pi r_0^4/4$ and $J=\pi r_0^4/2$ are the second moments of area for bending and twisting, respectively. 

\subsection{Fracture Criteria}
In the study by \cite{lu2021simulating}, the tensile and shear stresses within the bond in BDEM are calculated as:
\begin{align}
    \sigma &= \frac{\|\bm F_n\|}{S}+\frac{\|\bm M_b\| r_0}{I} \\
    \tau &= \frac{\|\bm F_s\|}{S}+\frac{\|\bm M_t\| r_0}{J},
\end{align}
where $\bm F_n$ and $\bm F_s$ represent the normal and shear forces acting on the bond, respectively. The quantities $|\bm M_b|$ and $|\bm M_t|$ indicate the magnitudes of the bending moments about the bond's normal and transverse axes, respectively, and $S$, $I$, and $J$ denote the cross-sectional area, second moment of area, and polar moment of area of the bond's cross-section, respectively. If the stress within the bond exceeds the respective strengths, i.e. $\sigma>\sigma_c$ or $\tau>\tau_c$ with $\sigma_c$ and $\tau_c$ representing the tensile and shear strengths, the bond breaks. 

In the explicit method, forces and torques are calculated directly, thus allowing the aforementioned fracture criteria to be readily computed. For our implicit method, where forces and torques are not computed directly, in order to employ the same fracture criteria, it is necessary to calculate the forces within the implicit setting. To achieve this, we have utilized the following equations:
\begin{align}
F_n &= \nabla_{p} V_{\mbox{stretch}}\\
F_s &= \nabla_{p} V_{\mbox{shear}} \\
\|M_b\| &= \|\langle K t, e_1 \rangle e_1 + \langle K t, e_2 \rangle e_2\| \\
\|M_t\| &= \|\langle K t, e_0 \rangle e_0\| ,
\end{align}
where $K$ and $t$ represent the stiffness matrix and rotation vector in the bending and twisting potential equation given by Eq. (\ref{eqn:bendtwist_potential}). In our experiments, we have found that this approach works effectively, yielding results that are nearly identical to those produced by the explicit method.

\subsection{Element Packing}
\label{sec:Element Packing}

In the domain of discrete element simulations, the concept of packing is analogous to the discretization process in continuum methods. Common packing techniques employed in discrete element simulations include random close packing and hexagonal close packing. For engineering applications that aim to accurately simulate material properties, random close packing combined with calibration is a frequently utilized approach. For the sake of simplicity in our application, we have employed hexagonal close packing. Our experiments have revealed that even with this most basic packing method, it is possible to effectively capture rich fracture details without introducing visually discernible artifacts that might result from overly regular patterns.

\begin{figure*}[ht]
    \centering
    \includegraphics[width=0.33\linewidth, trim=450 0 450 300, clip]{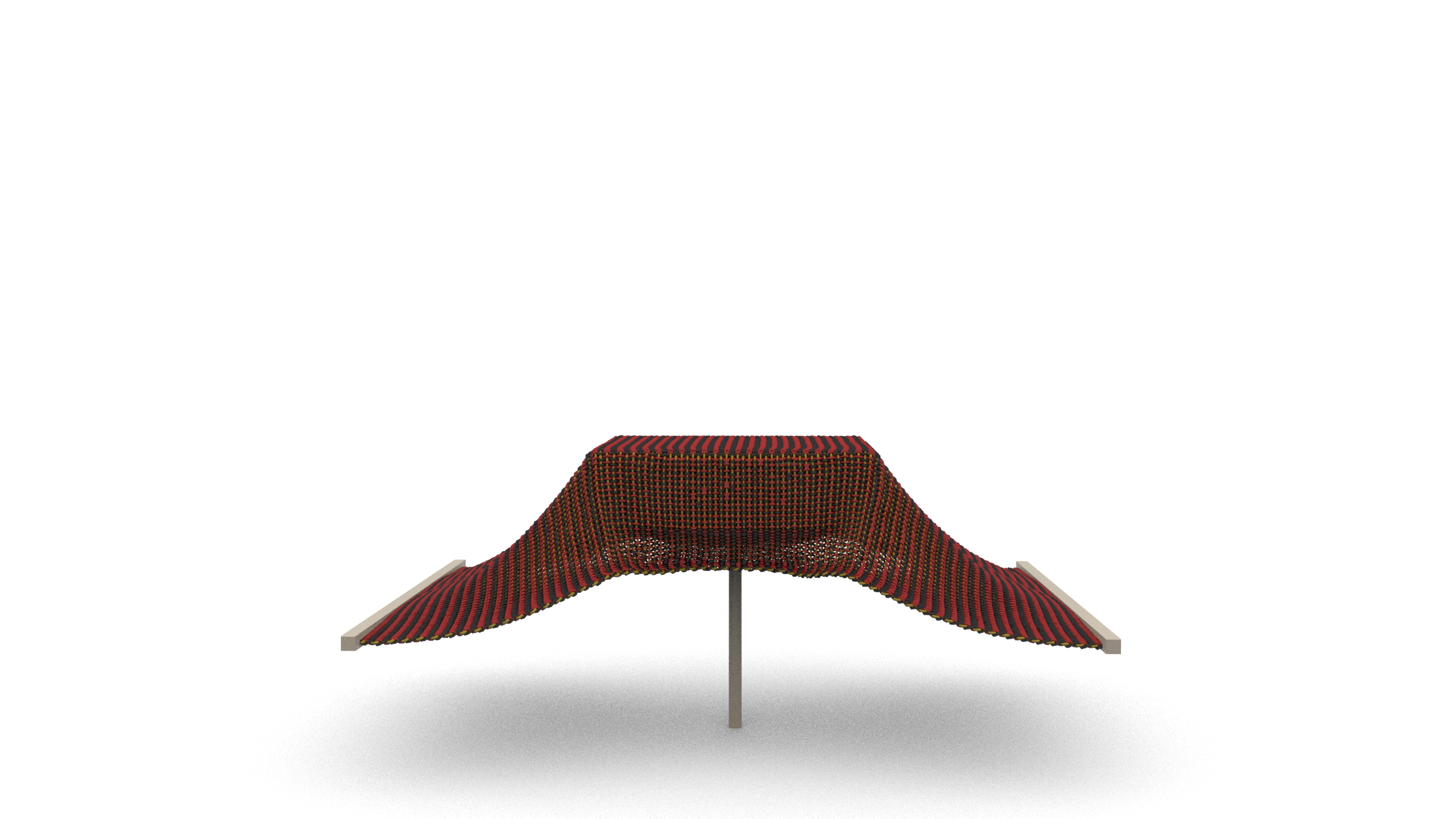}
    \includegraphics[width=0.33\linewidth, trim=450 0 450 300, clip]{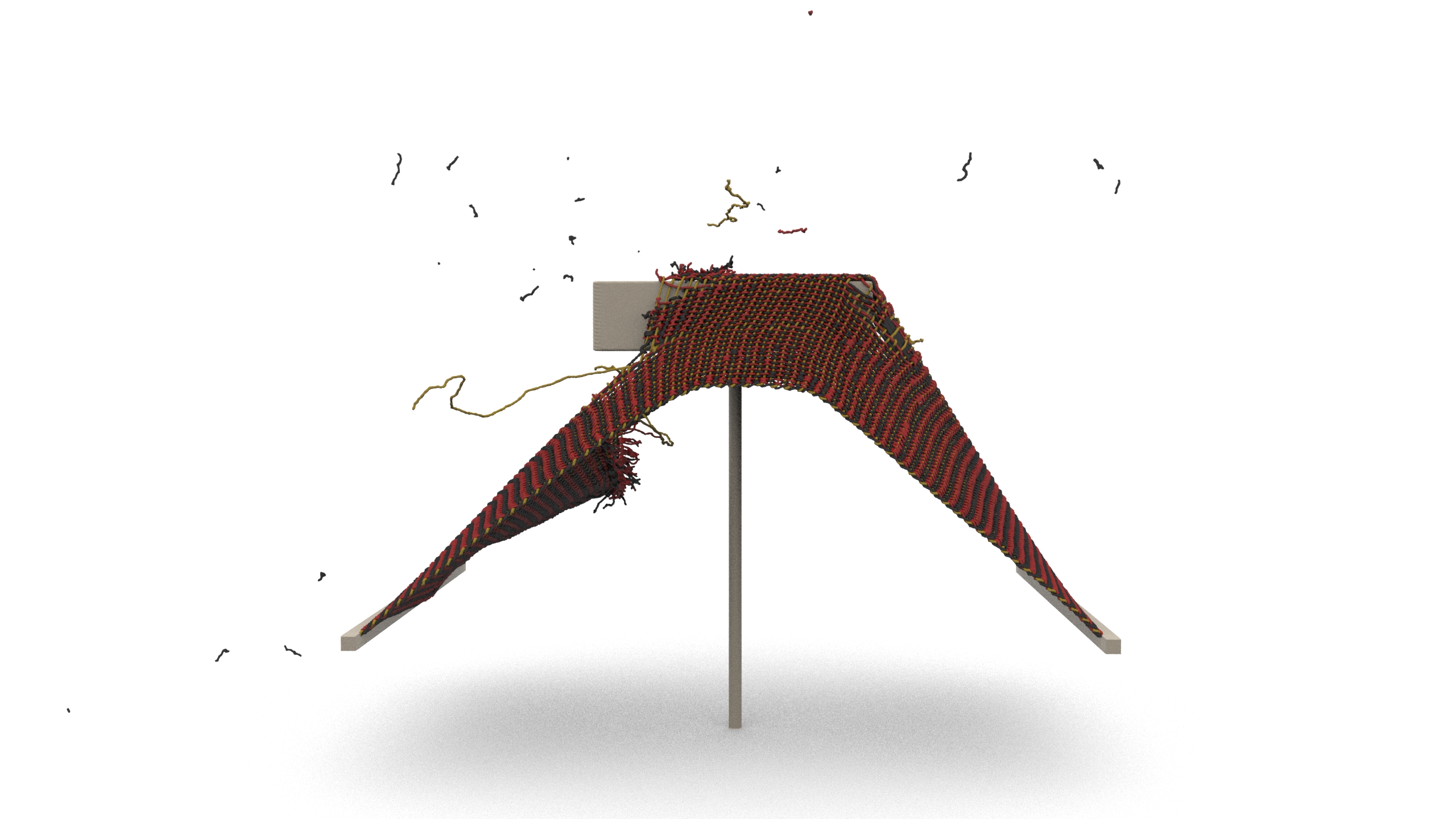}
    \includegraphics[width=0.33\linewidth, trim=450 0 450 300, clip]{figures/cloth.jpg}
    \caption{This scene illustrates the simulation of a woven fabric undergoing progressive stretching and eventual rupture using our method. With only 165K particles, our approach effectively captures the realistic behavior of yarn-level cloth and the phenomenon of fabric tearing, demonstrating a convincing rupture process and rich details at the tear propagation. Compared to current explicit methods, our technique achieves a threefold acceleration.}
    \label{fig:cloth}
\end{figure*}

\section{Manifold Optimization}
\label{sec:Manifold Optimization}

The above optimization approach generally improves the solution stability, but to solve it efficiently remains a challenge due to the quaternion constraint in Eq. (\ref{eqn:quaternion_constraint}). Our experiments have revealed that traditional methods such as the penalty method and the Lagrange multiplier method perform poorly. In this work, we present a manifold optimization scheme, where the quaternion constraint is represented as a hyper sphere enabling more effective optimization search on a manifold. Experimental results demonstrate that our new approach significantly improves the efficiency and accuracy of the BDEM solution compared to previous methods.

The manifold optimization approach \cite{boumal2022intromanifolds} replaces the constrained optimization problem on $\mathbb{R}^{n}$ with an unconstrained optimization problem on a specific manifold $\M$. In our approach, we utilize the second-order manifold optimization and employ a nullspace operator to further reduce the number of unknowns in the system. To better explain this solution process, we also introduce the first-order manifold optimization and compare its performance with our more advanced approach. To facilitate subsequent discussions, we establish specific nomenclature to differentiate functions defined on the full space from those on the manifold. The objective function is represented as $\originF$ in the original Euclidean space and as $f$ in the new manifold $\M$, such that $\originF|{\M} = f$. In our scenario, the manifold is defined as $\M = \prod_{i = 1}^{n} \mathbb{R}^{3} \otimes \mathbb{S}^{3}$, where $n$ denotes the number of particles. Additionally, $\grad f$ and $\Hess f$ refer to the gradient and Hessian in the manifold space, while $\grad \originF$ and $\Hess \originF$ pertain to the gradient and Hessian in the full space. The subscript $p$ denotes the $\mathbb{R}^{3}$ space, and the subscript $q$ denotes the $\mathbb{S}^{3}$ space.

\subsection{First-Order Manifold Optimization}
\label{subsec:First-Order Manifold Optimization}

\begin{algorithm}[htbp]
    \caption{First-Order Manifold Optimization}
    \label{alg:first_order_manifold_optimization}
    \begin{algorithmic}[1]
        \REQUIRE Initial guess $x_0$, termination threshold $\epsilon$
        \WHILE {true}
        \STATE Compute the $\grad f(x_k)$ on the manifold $\M$. \label{step:manifold_gradient}
        \IF {$\|\grad f(x_k)\| < \epsilon$}
        \RETURN $x_k \in \M$
        \ENDIF
        \STATE Use descent direction $-\grad f(x_k)$ with manifold line search step $\alpha$ to update $x_{k+1} \gets x_k - \alpha \grad f(x_k)$ on the manifold $\M$\label{step:manifold_linesearch}.
        \STATE Project $x_{k + 1}$ to the manifold $\M$.\label{step:manifold_project}
        \STATE $k \gets k + 1$
        \ENDWHILE
    \end{algorithmic}
\end{algorithm}

The first-order manifold optimization process is outlined in Algorithm \ref{alg:first_order_manifold_optimization}. It closely parallels standard gradient-based optimization methods but incorporates specific adjustments to align with the manifold. 

In Line \ref{step:manifold_gradient}, the gradient computation in manifold optimization involves a manifold tangent projection in comparison to standard gradient computation. Specifically, for our quaternion constraint, the gradient $\grad f$ is calculated as: 
\begin{align}
    \grad_{p} f &= \grad_{p} \originF \\
    \grad_{q} f &= (I - q q^{T}) \grad_{q} \originF ,
    \end{align}
where the subscript $p$ refers to the $\mathbb{R}^{3}$ space and the subscript $q$ refers to the $\mathbb{S}^{3}$ space. 

Line \ref{step:manifold_linesearch} is analogous to the line search and stepping procedures within standard gradient based optimization. In standard gradient based optimization, we employ direct addition to perform line searches and steps, which may lead to the system's state leaving the manifold. In contrast, manifold optimization requires that the corresponding line search and stepping be conducted on the manifold, ensuring that every step within the search and step progression adheres to the constraints of the manifold. Particularly, for our quaternion constraints, once the descent direction is known, we can ensure that the search and stepping remain on the manifold by employing a line search along a geodesic on $\mathbb{S}^{3}$. For element $i$ with position $p^i$ and rotation $q^i$, with a fast descent direction $\Delta p^i = -\grad^i_p f$, $\Delta q^i = -\grad^i_q f$ and step $\alpha$, the stepping on quaternion manifold is computed by
\begin{align}
    p^{i}_{\mbox{new}} &= p^{i} + \alpha\Delta{p^{i}} \label{eqn:manifold-update-1} \\
    q^{i}_{\mbox{new}} &= q^{i} \cos \|\alpha\Delta{q^{i}}\| + \frac{\alpha\Delta{q^{i}}}{\|\alpha\Delta{q^{i}}\|} \sin \|\alpha\Delta{q^{i}}\| \label{eqn:manifold-update-2} .
\end{align}
In the line search procedure, this approach is adopted for the incremental search of the system state. This same method is also applied during the final iterative update step.

In Line \ref{step:manifold_project}, the $q$ vector is guaranteed to lie on the manifold through a normalization step. This step can be regarded as a manifold projection, and is given by:
\begin{equation}
    q^i_k \gets \frac{q^i_k}{\|q^i_k\|} .
\end{equation}
It ensures that $q$ satisfies the manifold constraint, allowing the optimization process to proceed effectively.

\subsection{Second-Order Manifold Optimization}
\label{subsec:second_order_chap}
The second-order manifold optimization takes into account the Hessian and gradient of the objective function $f(x)$ on the manifold $\M$, and the algorithm steps are summarized in Algorithm \ref{alg:second_order_manifold_optimization}. 
\begin{algorithm}
\caption{Second-Order Manifold Optimization}
\label{alg:second_order_manifold_optimization}
\begin{algorithmic}[1]
    \REQUIRE Initial guess $x_0$, termination threshold $\epsilon$
    \WHILE {true}
    \STATE Compute the $\grad f(x_k)$ on the manifold $\M$. 
    \IF {$\|\grad f(x_k)\| < \epsilon$}
    \RETURN $x_k \in \M$
    \ENDIF
    \STATE Compute $\Hess f(x_k)$ on the manifold $\M$.\label{step:manifold_hessian}
    \STATE Solve $(\Hess f(x_k)) \Delta x = \grad f(x_k)$ in tangent space $\TM$.
    \STATE Use descent direction $\Delta x$ with manifold line search step $\alpha$ to update $x_{k+1} \gets x_k + \alpha \Delta x$ on the manifold $\M$.
    \STATE Project $x_{k + 1}$ to the manifold $\M$.
    \STATE $k \gets k + 1$
    \ENDWHILE
\end{algorithmic}
\end{algorithm}

The computation of $\Hess f$ in Line \ref{step:manifold_hessian} is performed using Proposition 5.8 in \cite{boumal2022intromanifolds}, and the procedure is summarized below: 
\begin{align}
    \Hess_{p_{i}, p_{j}} f &= \Hess_{p_{i}, p_{j}} \originF \label{eqn:manifold-hessian-1} \\
    \Hess_{p_{i}, q_{j}} f &= \Hess_{p_{i}, q_{j}} \originF (I - q_{j} q_{j}^{T})\label{eqn:manifold-hessian-2} \\
    \Hess_{q_{i}, p_{j}} f &= (I - q_{i} q_{i}^{T}) \Hess_{q_{i}, p_{j}} \originF \label{eqn:manifold-hessian-3} \\
    \Hess_{q_{i}, q_{j}} f &= (I - q_{i} q_{i}^{T}) \Hess_{q_{i}, q_{j}} \originF (I - q_{j} q_{j}^{T}) \\
    &\quad + \begin{cases}
        0 & i \neq j \\
        - q_{i}^{T} (\grad_{q_{i}} \originF) (I - q_{i} q_{i}^{T}) & i = j
    \end{cases} \label{eqn:manifold-hessian-4} .
\end{align}

The symmetric nature of $\Hess f$ allows us to solve the large sparse system using the commonly used preconditioned conjugate gradient method. With manifold optimization, the constrained optimization problem is transformed into an unconstrained problem. Even so, the unknowns of the system are still more than the true degrees of freedom of the physical system. To overcome this issue, we use the nullspace operator. Note that both the gradient project operator $P(q) = I - q q^T$ and the nullspace operator $G(q)$ have the similar property $P(q) q = 0$ and $G(q) q = 0$ (detailed introductions to these two operators are provided in ~\Cref{appendix:nullspace_operator}). Here we use $H = \Hess f$ and $b = \grad f$ as shorthand notation. The original second-order solution formulation for rotation $(I - q q^T) H (I - q q^T) \Delta x = (I - q q^T) b$ is reformulated as:
\begin{equation}
    G(q)^T G(q) H (G(q)^T G(q)) \Delta x = G(q)^T G(q) b .
\end{equation}
By defining $y = G(q) \Delta x$ and $z = G(q) b$, the original solution is transformed to:
\begin{equation}
    G(q) H G(q)^T y = z .
\end{equation}
The solution can then be expressed as $\Delta x = G(q)^T y$. This approach reduces the number of unknowns in the system to the DoFs of the BDEM system and improves the solution efficiency. 
%Our experiments reveal that this splitting demonstrates convergent properties equivalent to those of the original formulation.

\begin{table*}[ht]
    \centering
    \caption{The simulation statistics for our experiments are listed below, with the slow-motion of 1x corresponding to one frame per 1/60th of a second, and 0.125x corresponding to one frame per 1/480th of a second. The time step and time consumption for the proposed approach are compared with explicit and implicit methods at different element numbers and stiffness settings.}
    \begin{tabular}{ccccccccc}
    \thickhline
    Demo & $N$ & $E$ & $\Delta t$ (explicit) & $\Delta t$ (implicit) & Slow-Motion Scale & s/frame(explicit) & s/frame(implicit) & speedup \\ 
    \thickhline
    Beam Stretch & 19K & 1E7 & 2.6E-5 & 0.016 & 1x & 34.6 & 3.52 & 9.8 \\ \hline
    Beam Twist & 19K & 1E7 & 2.6E-5 & 0.0015 & 1x & 34.6 & 14.76 & 2.3 \\ \hline
    Beam Drape & 64K & 1E9 & 1.7E-6 & 0.008 & 1x & 2230.9 & 711.7 & 3.1 \\ \hline
    Beam Drape & 124K & 1E6 & 4.4E-5 & 0.0024 & 1x & 255.2 & 121.9 & 2.1 \\ \hline
    Chocolate & 58K & 1E9 & 4.4E-7 & 4.4e-5 & 0.125x & 857.8 & 343.8 & 2.5 \\ \hline
    Plate & 219K & 1E9 & 3.5E-7 & 3.5e-5 & 0.125x & 4564.6 & 1911.8 & 2.4 \\ \hline
    Cloth & 165K & 1E9 & 3.5E-7 & 1.74e-5 & 0.125x & 1299.6 & 394.2 & 3.3 \\
    \thickhline
    \end{tabular}
    \label{tab:params}
\end{table*}

\section{Implementation Details}
\label{sec:Implementation Details}

The utilization of optimization-based integrator formulation and manifold optimization helps to develop an unconditionally stable integrator suited for implicit BDEM simulation with large time steps. The simulation workflow is summarized in Algorithm \ref{alg:stable_solve}, and the implementation details are explained in the following subsections. The proposed methodology significantly improves the stability and efficiency of BDEM simulations, enabling them to tackle complex and computationally demanding scenes. 

\begin{algorithm}
    \caption{Stable Integrator}
    \label{alg:stable_solve}
    \begin{algorithmic}[1]
        \REQUIRE timestep $\Delta t$, termination threshold $\epsilon$
        \STATE $k \gets 0$ 
        \STATE Initial guess $x_0 \gets x_n + v_n \Delta t$
        \WHILE {true}
        \STATE Compute the $\grad f(x_k)$ on the manifold $\M$
        \IF {$\|\grad f(x_k)\| < \epsilon$}
            \STATE $x_{k+1} \gets x_k$
            \STATE \textbf{break}
        \ENDIF
        \STATE Compute $\Hess f(x_k)$ on the manifold $\M$
        \STATE Project $\Hess f(x_k)$ to semi-positive definite $\Hess^{'} f(x_k)$
        \STATE Compute relative error $\epsilon_r$
        \STATE Solve $(\Hess^{'} f(x_k)) \Delta x = -\grad f(x_k)$ in tangent space $\TM$ with preconditioned conjugate gradient method with relative error $\epsilon_r$
        \STATE Use descent direction $\Delta x$ with manifold line search step $\alpha$ to update $x_{k+1} \gets x_k + \alpha \Delta x$ on the manifold $\M$.
        \STATE Project $x_{k + 1}$ to the manifold $\M$.
        \STATE $k \gets k + 1$
        \ENDWHILE
        \STATE $x_{n+1} \gets x_{k + 1}$.
        \STATE $v_{n+1} \gets \frac{1}{\Delta t}(x_{n + 1}-x_n)$
        \STATE Project $v_{n+1}$ to friction cone
        \STATE Update bond state with $x_{n+1}$
    \end{algorithmic}
    \end{algorithm}

\subsection{Timestep Selection}

In our research, the selection of time steps for both implicit and explicit simulations requires careful consideration. For explicit simulations, we must choose a time step that is sufficiently large to permit a reasonable comparison of efficiency. Regarding implicit simulations, the time step should be sufficiently increased to ensure significant performance acceleration without excessively compromising the realism of the fragmentation effects. Taking into account these factors, we have followed the approach established in explicit BDEM for estimating the system's period to inform our choice of an appropriate time step. In \cite{lu2021simulating}, the appropriate time step is ascertained by analyzing the system's period $T$, which correlates with the system's largest eigenvalue $\lambda$ as delineated by the equation $T = 2 / \sqrt{\lambda}$. The largest eigenvalue $\lambda$ of the system is calculated using the expression
\begin{equation}
    \lambda = \frac{k}{m}N,
\end{equation}
where $m$ denotes the effective mass, $k$ signifies the stiffness coefficient, and $N$ corresponds to the maximum eigenvalue of the Laplacian matrix representing the inter-bond connections. For a hexagonal arrangement, $N\approx 16$. Additionally, the stiffness $k$ is given by $k = k_n \approx \frac{\pi}{2}E r$, with $E$ representing the Young's modulus and $r$ being the average radius of the elements. 

For explicit simulation, our experimental findings indicate that a time step of $0.2T$ is the threshold that avoids numerical instability and oscillations. This time step is consistently applied across all explicit simulation comparisons with implicit methods presented in the text. Regarding implicit simulations, our optimization-based method supports stable computations with large time steps, potentially allowing for one time step per frame. However, increasing the time step can lead to more iterations in the Newton's solver, decreasing search efficiency. Beyond a certain point, the computational savings from increasing the time step are not substantial. Additionally, an overly large time step may result in infrequent fracture detection, which can impair the simulation's accuracy. In our experiments, we have found that a time step ranging from 20 to 100 times the period $T$ is reasonable.

\subsection{Semi-Positive Definite Projection}
The Hessian matrix on a manifold may not be semi-positive definite, which poses a significant challenge as it may lead to a wrong search direction. To guarantee that the search direction is a correct descent direction, a semi-positive definite projection onto the manifold is applied to our Hessian matrix. Similar regularization treatments are also found in other methods \cite{li2020incremental}. 

\subsection{Preconditioned Conjugate Gradient}
The preconditioned conjugate gradient method is employed to efficiently solve large sparse symmetric systems of the form $\Hess f(x_k)$. Specifically, we employ an algebraic multigrid (AMG) precondition to facilitate fast and accurate solutions. Similar to \cite{gast2015optimization}, the absolute error is not explicitly managed in the beginning steps, and instead the relative error is taken as a termination criterion for the conjugate gradient method. This relative error is calculated as:
\begin{equation}
    \epsilon_r = \min(0.5, \sqrt{|\grad f(x_k)|}).
\end{equation}
The AMG preconditioning in conjunction with the relative error termination criterion lead to efficient and accurate solutions for large sparse symmetric systems.

\begin{figure*}[ht]
    \centering
    \includegraphics[width=0.25\linewidth]{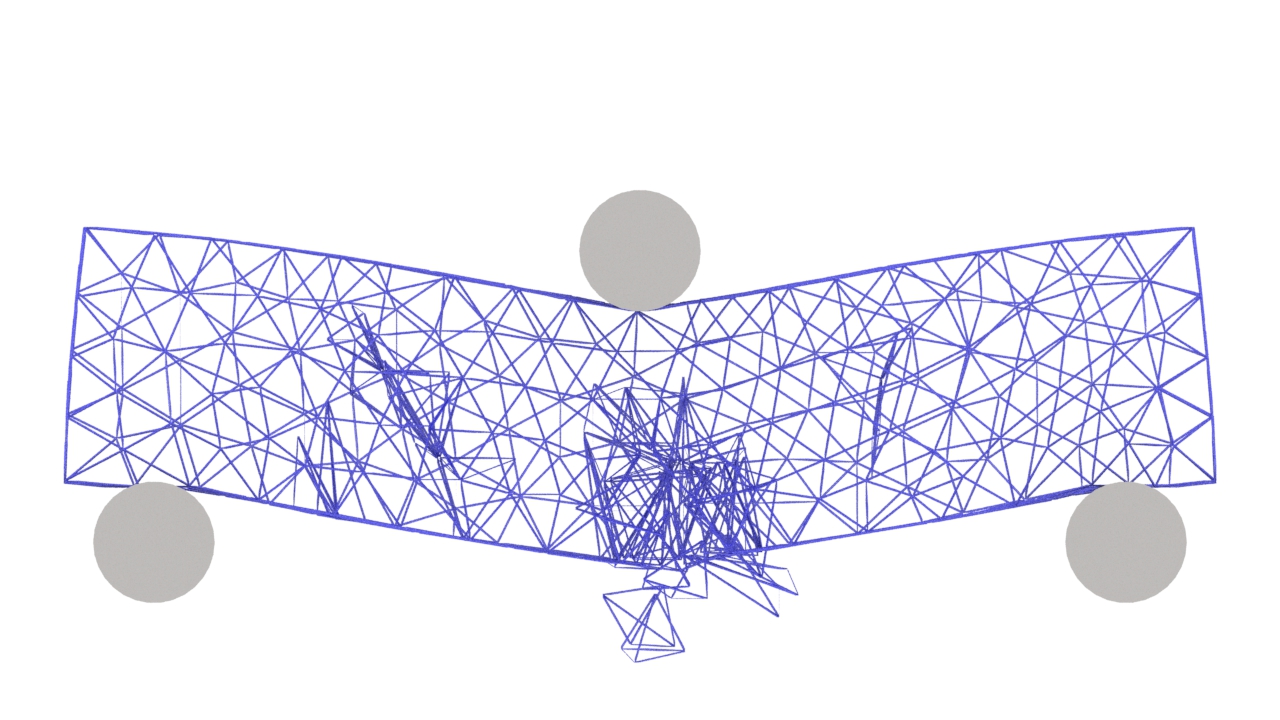}
    \hspace{10mm}
    \includegraphics[width=0.25\linewidth]{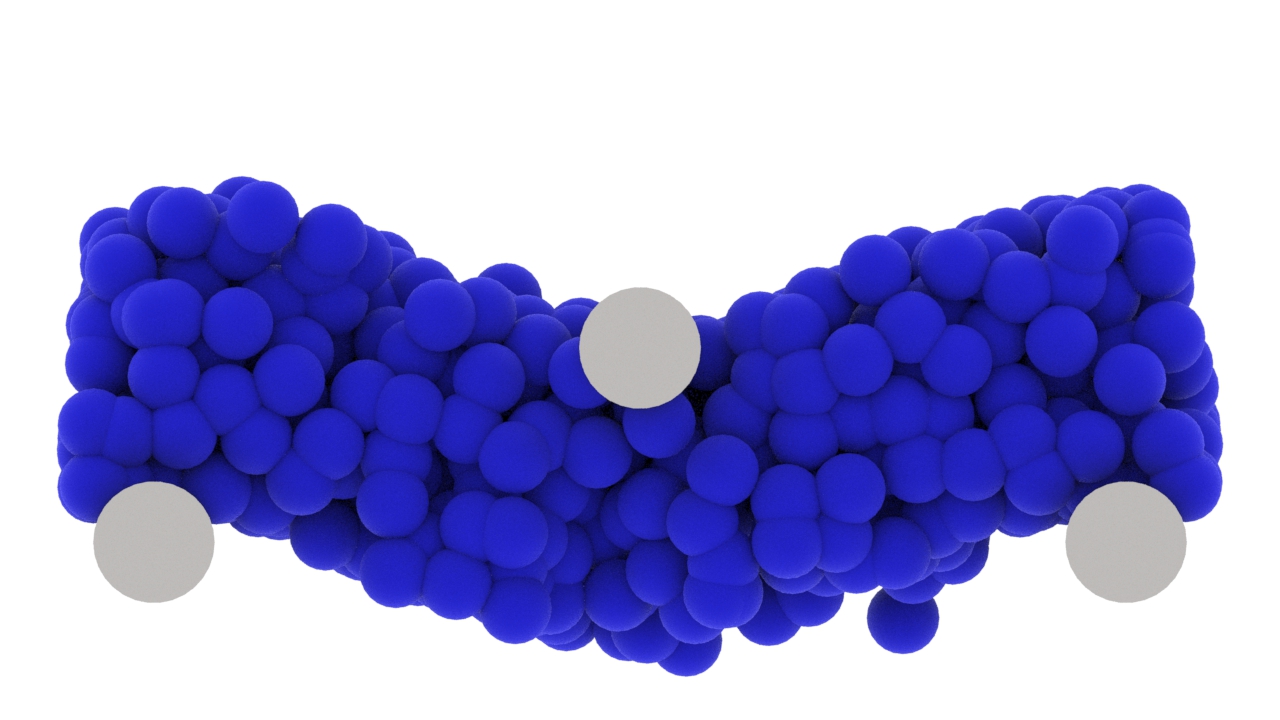}
    \hspace{10mm}
    \includegraphics[width=0.25\linewidth]{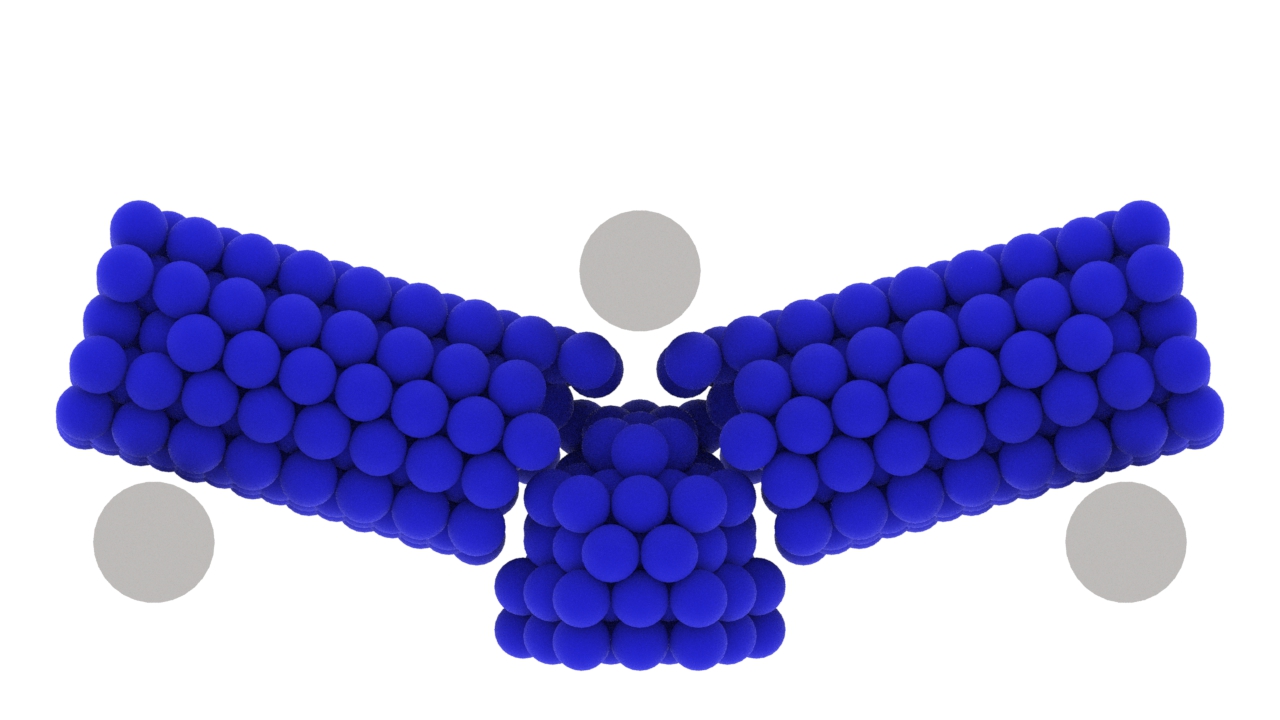}
    
    \includegraphics[width=0.25\linewidth]{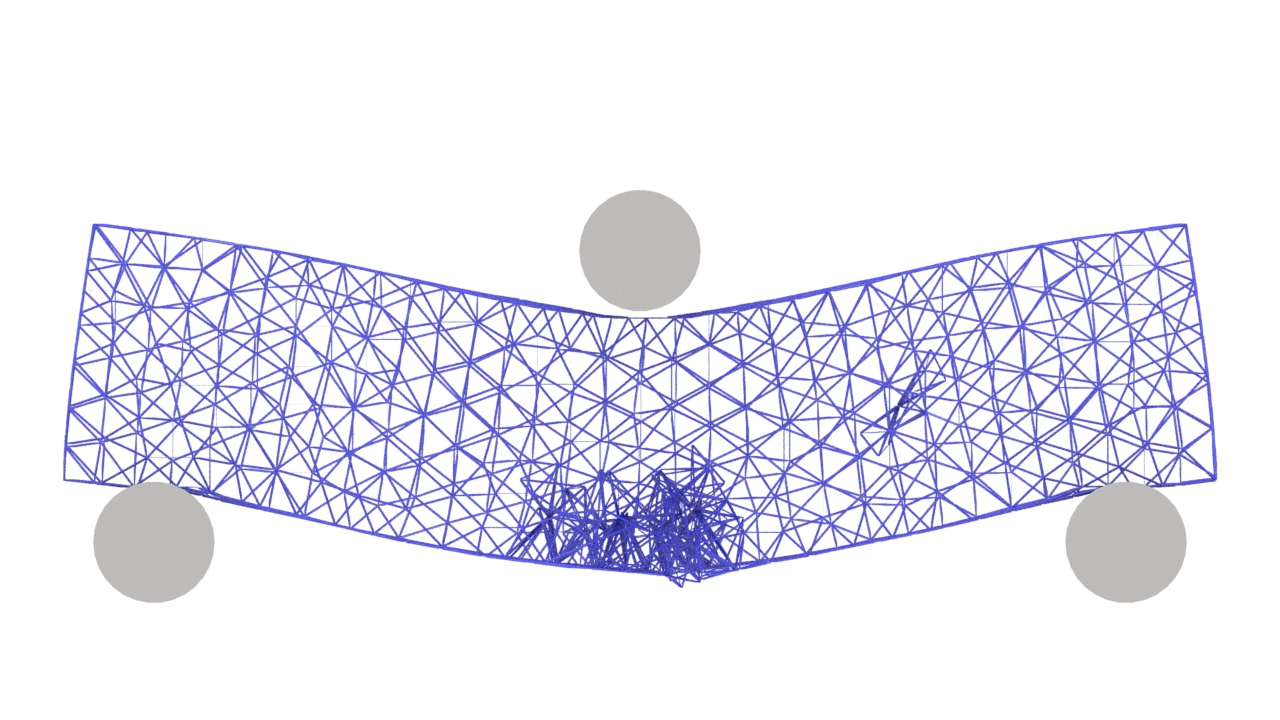}
    \hspace{10mm}
    \includegraphics[width=0.25\linewidth]{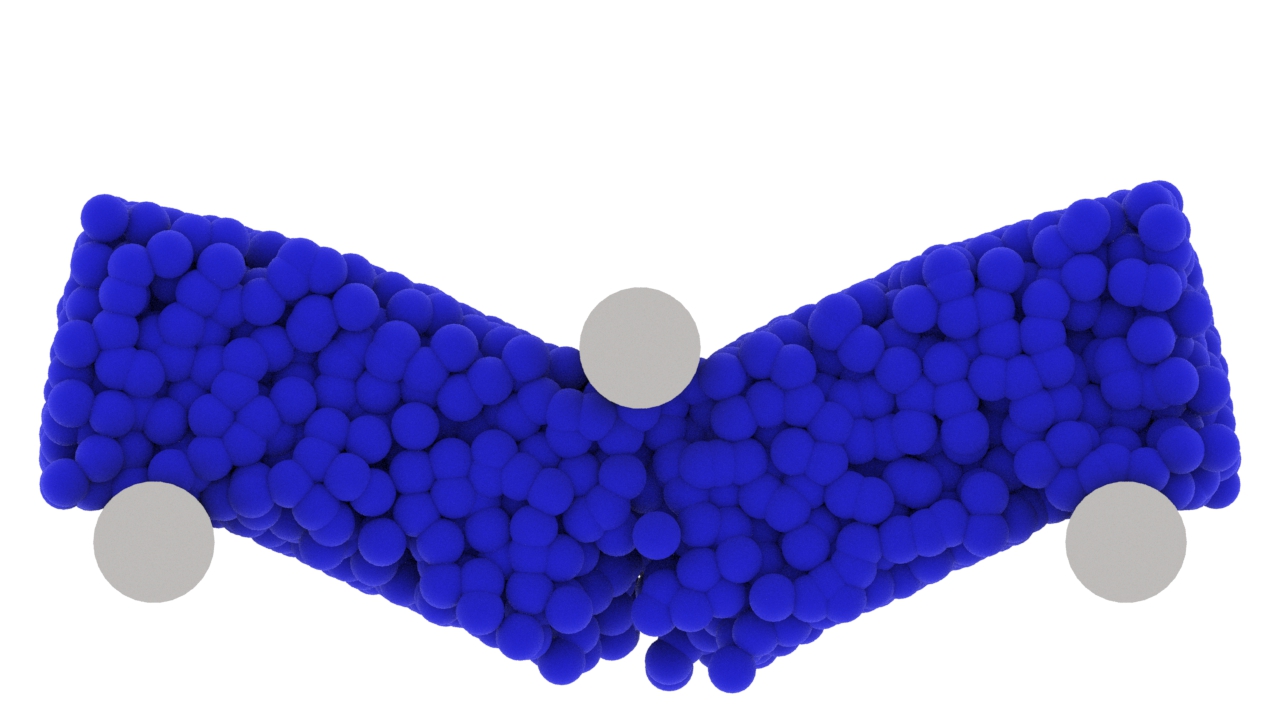}
    \hspace{10mm}
    \includegraphics[width=0.25\linewidth]{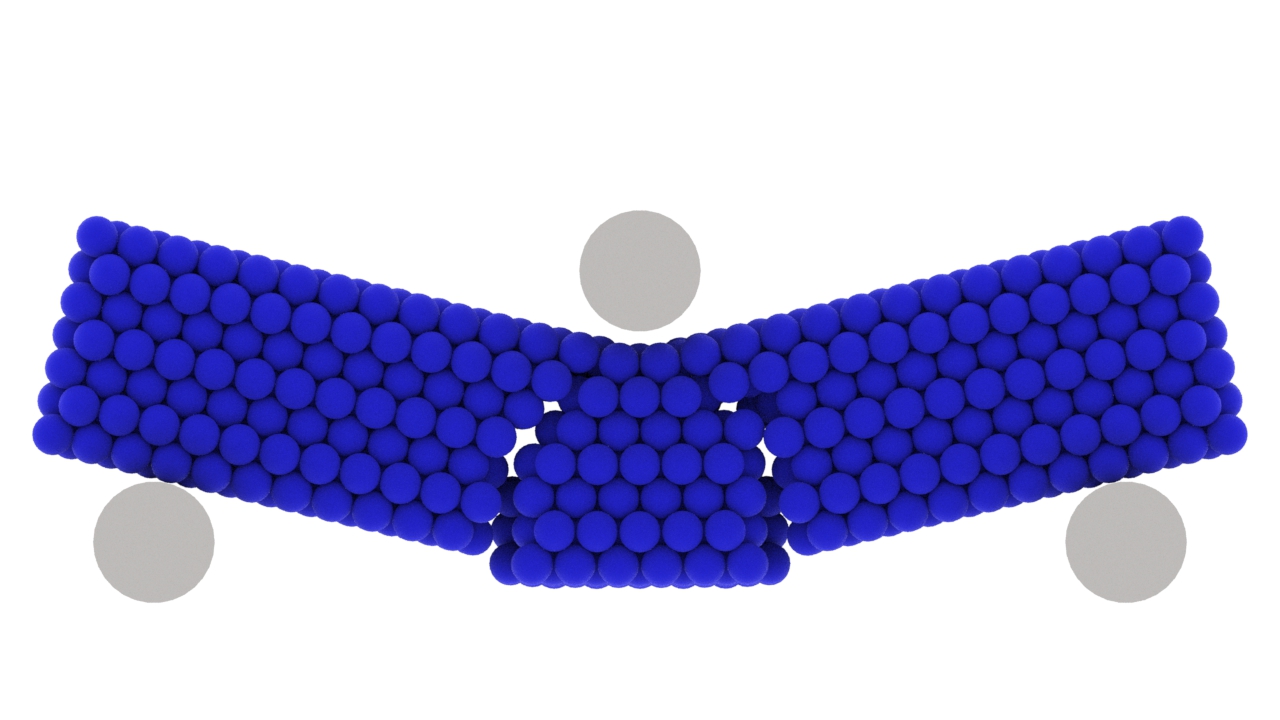}
    
    \includegraphics[width=0.25\linewidth]{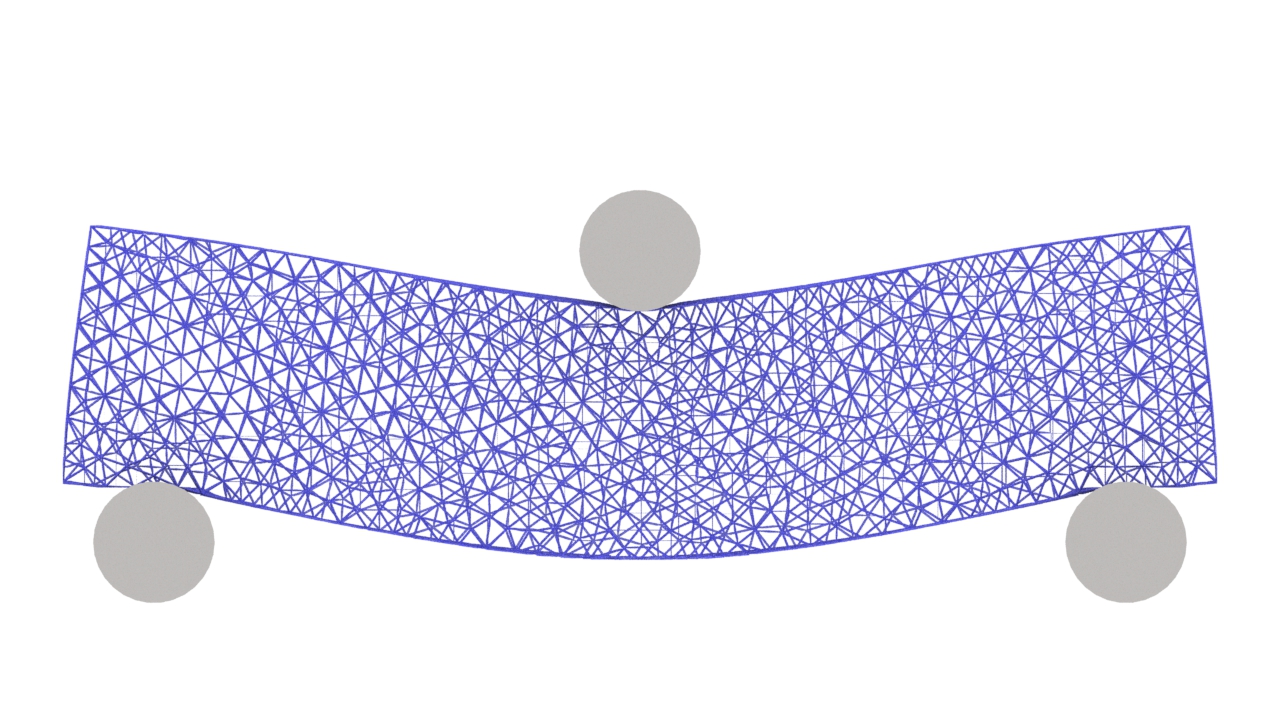}
    \hspace{10mm}
    \includegraphics[width=0.25\linewidth]{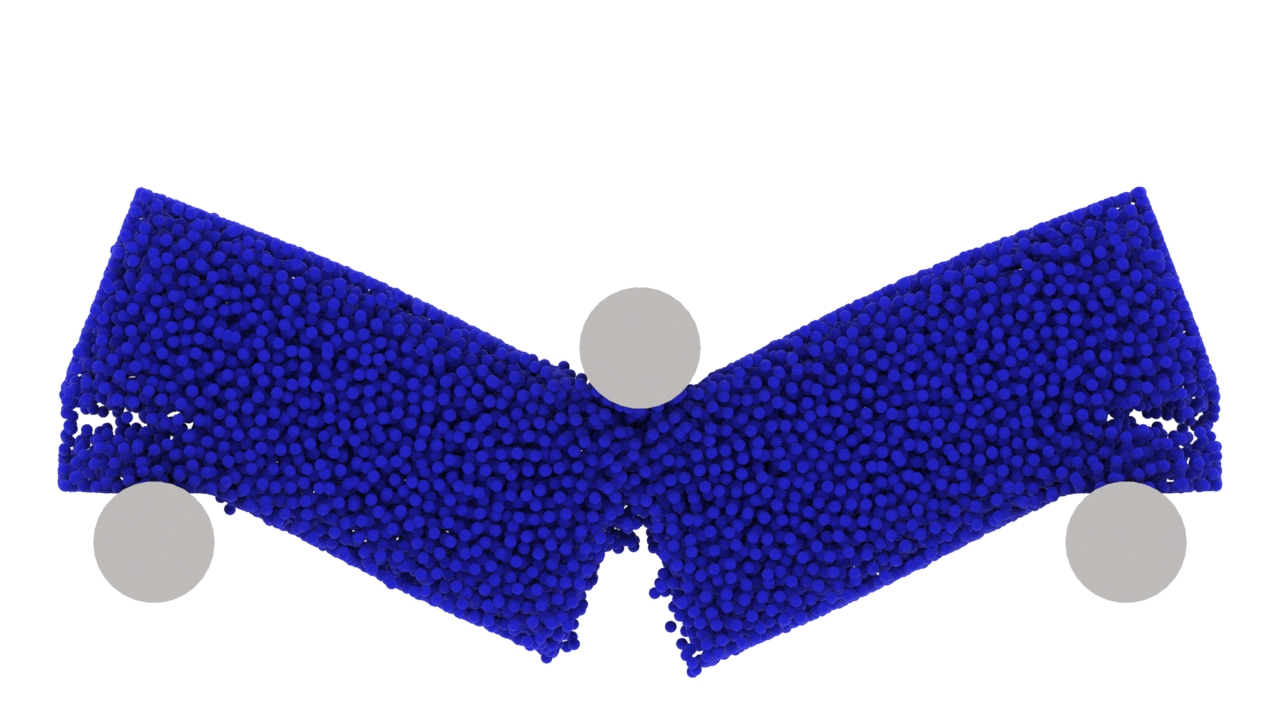}
    \hspace{10mm}
    \includegraphics[width=0.25\linewidth]{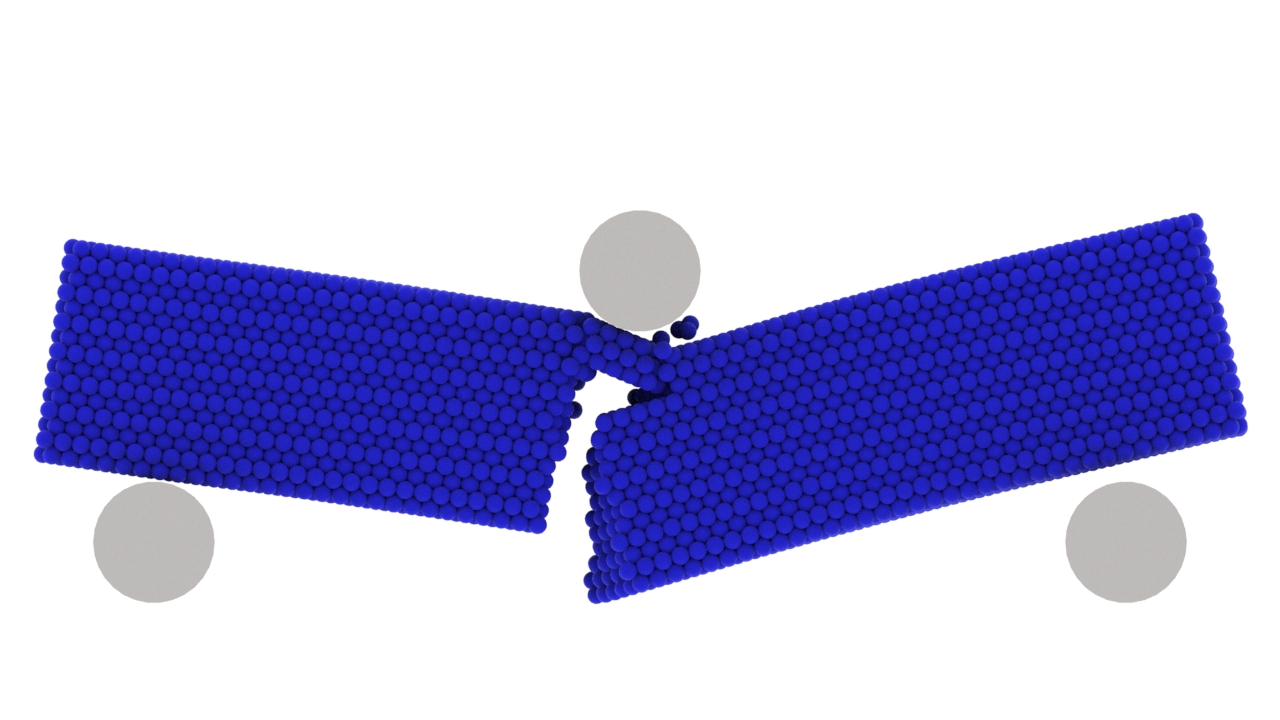}
    \caption{The figure, from left to right, illustrates the performance of the FEM, MPM, and our approach across varying scales. The upper, middle, and lower rows represent the fracture behavior of the three methods under identical loads at different scales. It is observable that FEM exhibits completely distinct fracture occurrences across the three scales. MPM shows similar fracture timings at medium and high scales but deviates at lower scales. In contrast, our method consistently demonstrates fracture at the same load across all scales, indicating superior scale consistency.}
    \label{fig:three_point_bending}
\end{figure*}

\subsection{Collision Handling}

In accordance with our optimization formulation, a variety of collision handling methods can be seamlessly integrated into our approach, including simple penalty-based methods and recently utilized incremental potential contact \cite{li2020incremental}. However, this paper primarily focuses on fragmentation-related phenomena, where the impact of the collision module on the fragmentation performance is minimal and not the main emphasis here. Therefore, we have employed a very simple treatment method for visually acceptable simulation results.

\paragraph{Repulsion Term}
A simple collision potential is introduced to model repulsion between discrete elements and external objects. Specifically, the potential includes two components: $V_{\mbox{self}}$, representing the repulsive forces between the discrete elements, and $V_{\mbox{external}}$, representing the repulsive forces between the elements and the external objects. The corresponding mathematical expressions are:
\begin{align}
    V_{\mbox{self}} &= \frac{1}{2}k_{ec}(\|p_i - p_j\| - r_i - r_j)^2 \\
    V_{\mbox{external}} &= \frac{1}{2}k_c\|g(p_i) \nabla g(p_i)\|^2,
    \end{align}
where $k_{ec}$ denotes element collision stiffness, $k_c$ external collision stiffness, $p_i$ and $p_j$ the positions of the discrete elements, and $g$ the signed distance function of the external objects. In order to ensure that the collision term does not adversely affect the system's computational efficiency while maintaining a visually coherent outcome without noticeable interpenetration, we have set the collision stiffness and element collision stiffness to the order of $E r_\text{max}$, where $E$ is the max Young's Modulus of element and $r_\text{max}$ is the max radius of elements. In our experiments, this selection has minimal impact on the system's convergence rate and yields visually acceptable simulation results.

\paragraph{Friction}
Our approach for modifying the velocity and angular velocity of an element in the post-collision process relies on an explicit method. In the case of an external collision or self-collision, the relative velocity $v_{ij}$ and the relative angular velocity $w_{ij}$ are computed between element $i$ and the external object $j$ or another element $j$, respectively, using $v_{ij} = v_n + v_t$, where $v_n$ and $v_t$ denote the normal and tangential components of the relative velocity. To calculate the modification of velocity and angular velocity, we use the following equations:
\begin{align}
\Delta v_i &= -\min(1.0, \mu_s \frac{\|F_n\|\Delta t}{m_i \|v_t\|}) v_t \\
\Delta w_i &= -\min(1.0, \mu_r \frac{\|F_n\|r_i^3 \Delta t}{I_i \|w_{ij}\|}) w_{ij} .
\end{align}
Here, $m_i$ and $I_i$ represent the mass and moment of inertia of element $i$, respectively, and $\mu_s$ and $\mu_r$ denote the coefficients of sliding and rotation friction. For external collisions, $F_n = k_c g(p_i) \nabla g(p_i)$, and for self-collisions, $F_n = k_{ec}|\|p_i - p_j\| - r_i - r_j|\frac{p_i - p_j}{\|p_i - p_j\|}$. Finally, we project the modifications onto the direction of $v_i$ and $w_i$, and ensure that the magnitude of the modification does not exceed the modulus of $v_i$ and $w_i$. Therefore, the velocity and angular velocity can be modified to zero at most. The explicit friction treatment method is evidently not sufficiently precise. In our experiments, we observed that different time step selections necessitated the use of varying friction coefficients to achieve similar visual outcomes. Smaller time steps required a reduction in the friction coefficient, while larger time steps demanded an increase. However, since the friction effect is not the focus of this paper, we intend to employ a more accurate model in future work to obviate the need for such parameter adjustments.

\section{Results and Discussion}
\label{sec:Results and Discussion}

The performance of the proposed approach is evaluated through a series of experiments. The simulations were conducted using an Intel i7 13700K processor, and detailed experimental setups and time costs are presented in Table \ref{tab:params}. In all experiments, we have employed the conventional particle fluid surface method for the reconstruction and rendering of fracture surfaces.

\subsection{Comparison with Rotation Vector Representation}

\begin{figure}[htbp]
    \centering
    \includegraphics[width=\linewidth]{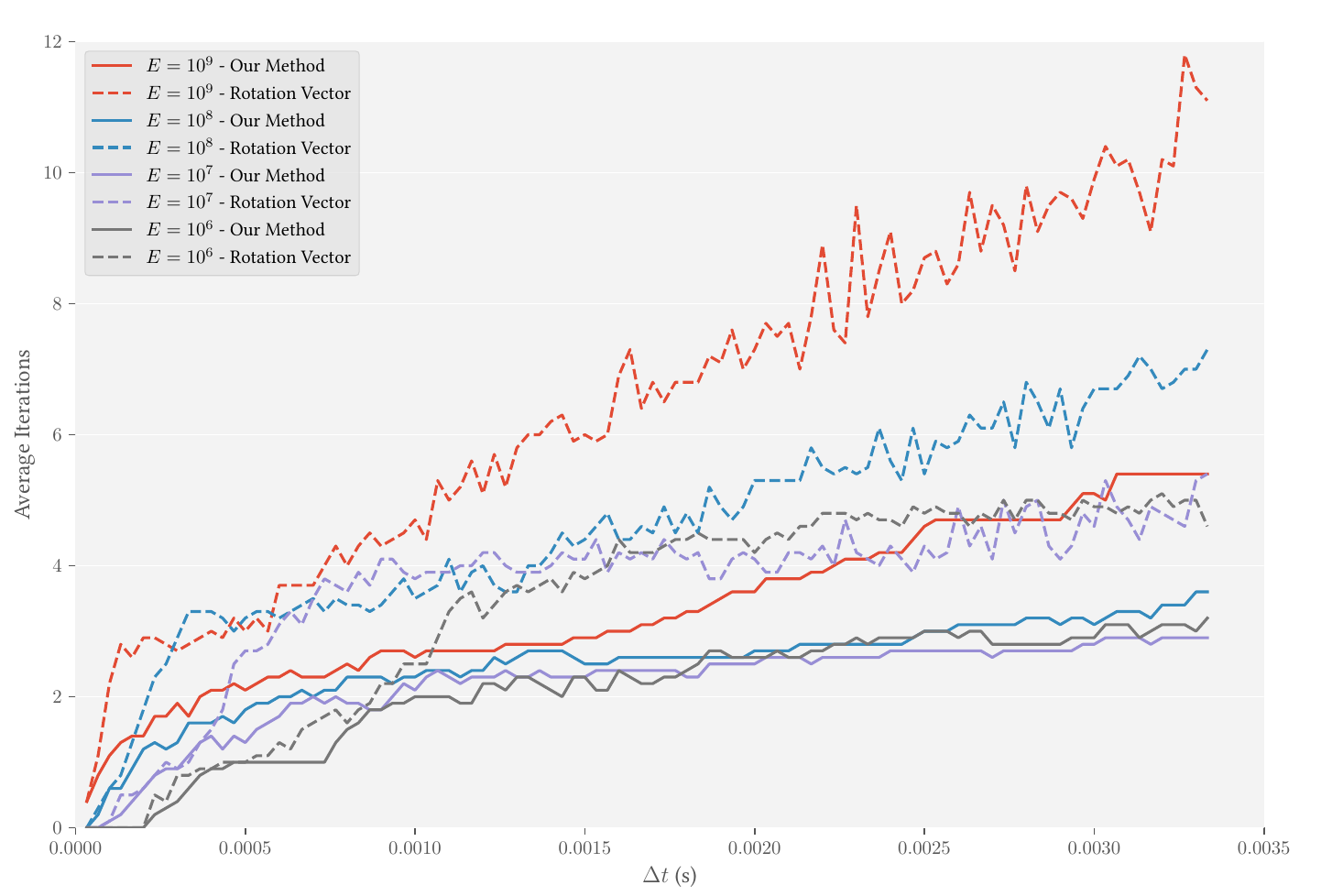}
    \caption{We conducted a three-point beam bending experiment with identical settings for our method and rotation-vector formulation in \cite{geilinger2020add}. The horizontal axis represents the time step size used in the simulation, with curves of different colors indicating various material Young's moduli. The vertical axis denotes the average number of iterations required during the simulation process. The dashed lines in the figure represent the rotation vector formulation, while the solid lines represent our approach. It is observable that, across different time steps and stiffness coefficients, our method needs fewer iterations to achieve the same convergence criteria.}
    \label{fig:compare_add}
\end{figure}

In the field of rigid body motion simulation, the rotation vector representation is commonly employed to compute the rotational degrees of freedom. Compared to our method, the rotation vector also upholds a computational size of six degrees of freedom. We have conducted experimental comparisons between our motion representation and the rotation vector-based motion formulation from ADD \cite{geilinger2020add}. Detailed experimental results are presented in Figure \ref{fig:compare_add}. Utilizing the same BDEM Potential, we configured an identical three-point beam bending experiment and set an equivalent dimensionless convergence threshold. Our experiments revealed that the simulation outcomes were indistinguishable between the two methods; however, our algorithm achieved faster convergence than the ADD approach. Theoretically, the rotation vector representation introduces no constraints, suggesting that our manifold-based search algorithm should exhibit a similar convergence rate as the rotation vector. We preliminarily attribute the enhanced performance of our algorithm to two factors: firstly, the quaternion representation, which offers superior spatial interpolation properties favorable for search, and secondly, unlike our energy minimization-based method, the ADD approach requires the application of the Gauss-Newton method to solve the motion equations, which may impose limitations on the convergence efficiency to a certain extent.

\subsection{Comparison with Other Optimization Methods}
\begin{figure}[htbp]
\centering
\includegraphics[width=\linewidth]{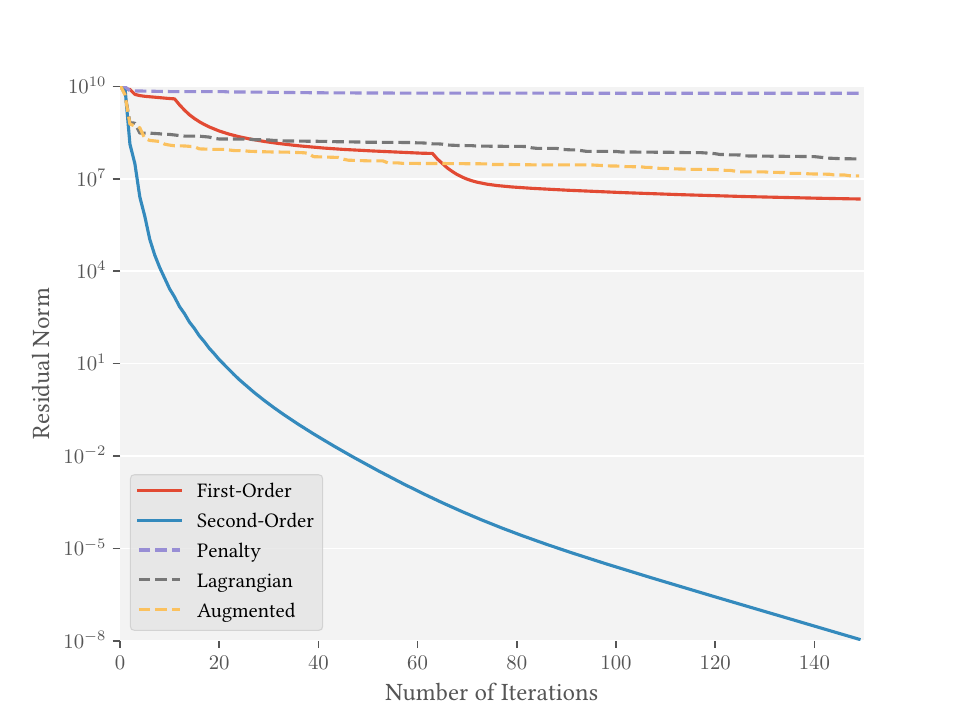}
\caption{Convergence rate comparison. The results show that the first-order manifold optimization and the second-order manifold optimization are both faster than the conventional constrained optimization methods, including the penalty method and the Lagrange multiplier method.}
\label{fig:optimization}
\end{figure}

To examine the performance of our proposed optimization approach, several experiments were conducted to assess its outcomes against commonly used constrained optimization methods. The results of these experiments are presented in Figure \ref{fig:optimization}, which shows our proposed approach outperforms commonly used constrained optimization methods. 

\paragraph{Penalty Method}
The penalty method uses an external potential to convert a constrained optimization problem to an unconstrained optimization problem:  
\begin{equation}
    \label{eqn:penalty_method}
    \argmin_{\bm x} \Phi(\bm x) + \frac{1}{2} \kappa \| C(\bm x) \|^2 .
\end{equation}
This method is commonly used in constrained optimization due to its simplicity. However, its shortcoming is that indefinite stiffness is required to satisfy constraints, and the introduced external stiffness $\kappa$ can make the entire system an ill-defined problem. In our experiments, we found that the penalty method failed to converge or converged very slowly in BDEM fracture simulations. 

\paragraph{Lagrange Multiplier}
The Lagrange multiplier adds external variables $\lambda$ to the system to convert the constrained optimization problem to an unconstrained one: 
\begin{align}
    \label{eqn:lagrangian_multiplier}
    \argmin_{\bm x, \lambda} \Phi(\bm x) + \lambda^T C(x) .
\end{align}
One drawback of this method is the increase in the number of variables and computational cost. The actual DoFs of rotation is $3$, but the use of quaternions increases the number of unknowns to $4$. The addition of multipliers further increases the number of variables to $5$. As the number of unknowns increases, the computational cost also increases. In our experiments, we found that the convergence rate was even slower than that of the first-order manifold optimizer. 

\paragraph{Augmented Lagrangian}
The augmented Lagrangian method can be seen as a combination of the penalty method and the Lagrange multiplier method, and it turns a constrained optimization problem into an unconstrained one. 
\begin{align}
    \argmin_{\bm x, \lambda} \Phi(\bm x) + \lambda^T C(x) + \frac{1}{2} \kappa \| C(\bm x) \|^2 .
\end{align}
The external stiffness does not need to be infinite for the augmented Lagrangian method. However, in our experiments, the convergence rate was still lower than that of the manifold optimization method.

\begin{figure*}[ht]
    \centering
    \includegraphics[width=0.33\linewidth, trim=200 0 200 0, clip]{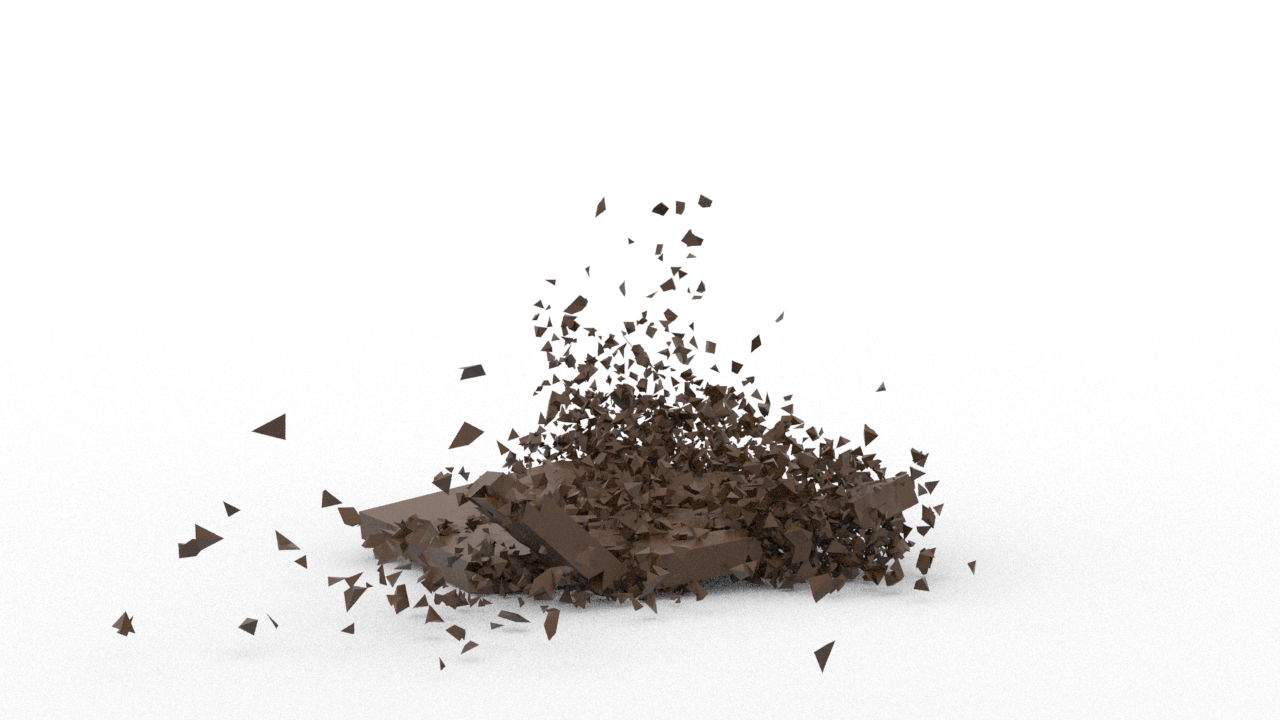}
    \includegraphics[width=0.33\linewidth, trim=200 0 200 0, clip]{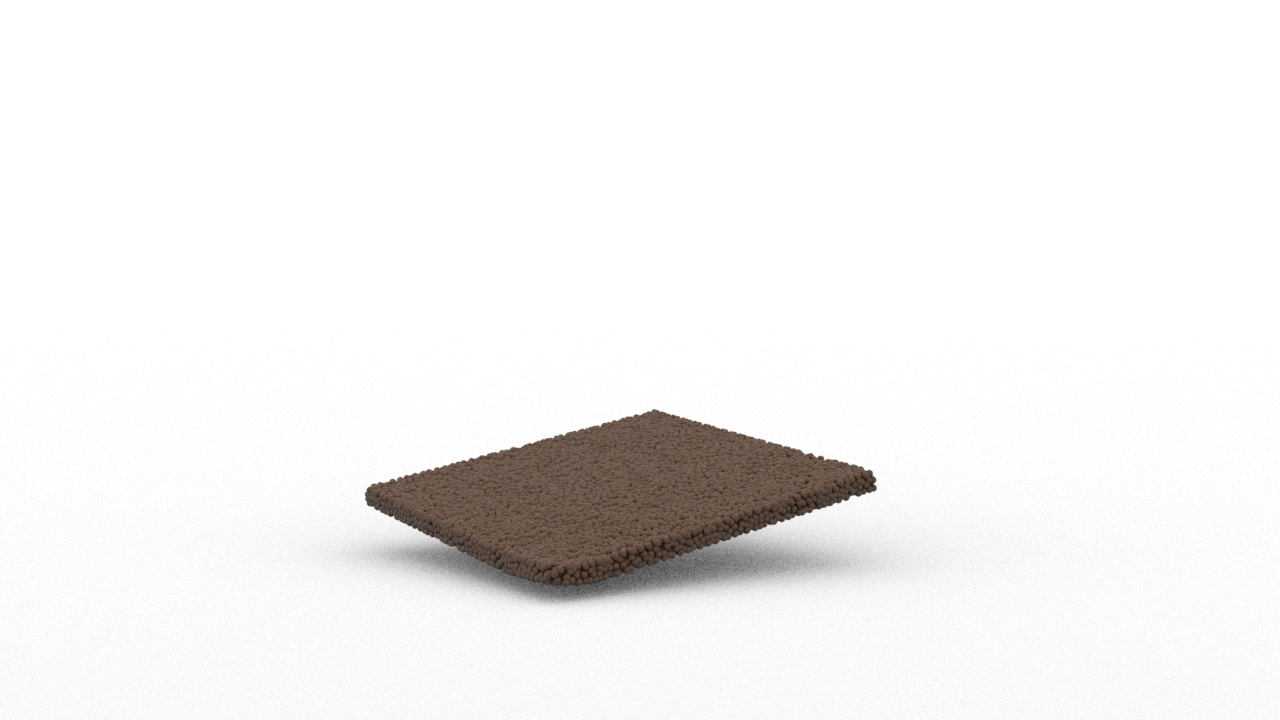}
    \includegraphics[width=0.33\linewidth, trim=200 0 200 0, clip]{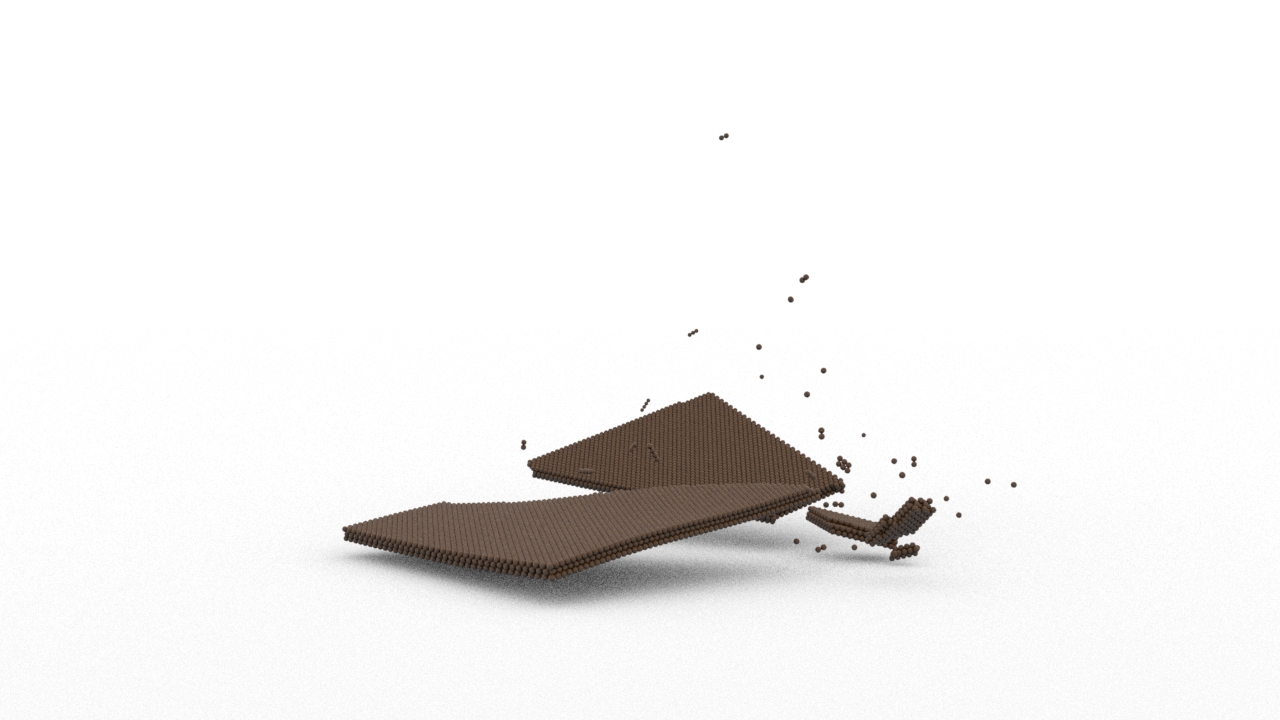}
    \caption{This figure, from left to right, demonstrates the fracture performance of FEM, MPM, and our method under the high-speed drop impact of a chocolate object. The FEM resulted in an excessive fragmentation into minute pieces, while the MPM failed to produce a physically plausible fragmentation. In contrast, BDEM accurately fragmented into a variety of piece sizes, effectively capturing the desired material failure behavior.}
    \label{fig:chocolate_compare}
\end{figure*}

\subsection{Comparison with FEM and MPM}
FEM and MPM are two prevalent techniques for simulating fragmentation. FEM has been utilized in real-time simulations \cite{o1999graphical,parker2009real}, while MPM can simulate fracture through the continuum damage model \cite{wolper2019cd,fan2022simulating}. We have implemented the methodologies presented in \cite{parker2009real} and \cite{wolper2019cd} as benchmarks for FEM and MPM, respectively, and conducted two experiments for a detailed comparison: the three-point bending test and the chocolate falling test. The three-point bending experiment highlights the differences in scale consistency among the three methods across various scales, and the chocolate falling test illustrates the disparities of fragmentation under high-velocity impacts.

To ensure a fair and comprehensive comparison, we provide implementation details of the baseline methods used in our experiments. For the MPM-based approach, we utilize the open-source phase field fracture model presented in \cite{wolper2019cd}. Our FEM implementation is based on the method described in \cite{parker2009real}, with several modifications to enhance accuracy while maintaining computational efficiency. We reintroduce the residual propagation and sequential separation updating techniques from \cite{o1999graphical} to achieve more precise fracture propagation. We remove the upper limit on propagation steps, allowing for a more faithful representation of the fracture process. Additionally, we eliminate the minimum fracture piece size constraint of three elements to fully capture the algorithm's behavior. To maintain physical plausibility, we implement an additional check to remove connections with floppy hinge or ball joints, as suggested in \cite{parker2009real}. This modified FEM implementation can be viewed as a simplified version of \cite{o1999graphical}, with the exclusion of the remeshing process for individual tetrahedra. This strategic simplification prevents an increase in tetrahedron count during simulation while still capturing key aspects of fracture behavior. The resulting baseline offers a balance between computational efficiency and physical fidelity.

\paragraph{Three Point Bending}

Scale consistency is crucial for artists during the rapid prototyping phase of simulation, as it allows them to experiment with parameters at a smaller scale, thereby avoiding the time-consuming iteration associated with large-scale simulations. To analyze the scale consistency of our method in comparison to FEM and MPM, we conducted a three-point bending experiment. This test is a standard for analyzing material properties, where the relationship between beam fracture and the applied load reveals the intrinsic characteristics of the material. As depicted in the \Cref{fig:three_point_bending}, we selected discrete representations of the same material for the three methods across three different scales. FEM exhibits markedly different fracture events across scales. MPM demonstrates similar fracture behavior at medium and high scales but fails to accurately simulate fracture at smaller scales, with varying fracture initiation frames across resolutions (see \Cref{tab:comparison_params}). In contrast, our method maintains consistent fracture characteristics across all scales, demonstrating superior scale consistency. This consistency in our method arises from the discrete nature of fracture phenomena, clearly evident in the comparison experiment. Continuum approaches like FEM and MPM often struggle to accurately model small crack patterns due to insufficient local discretization, leading to resolution-dependent fracture initiation times. As fracture originates from small cracks, these continuum approaches may fail to capture the full complexity of the fracture process, resulting in resolution-dependent crack patterns. Our discrete method, however, inherently captures these small-scale interactions, yielding more consistent behavior across different scales.

\begin{figure}[htbp]
    \includegraphics[width=0.49\linewidth, trim=300 0 300 0, clip]{figures/chocolate_mpm.jpg}
    \includegraphics[width=0.49\linewidth, trim=300 0 300 0, clip]{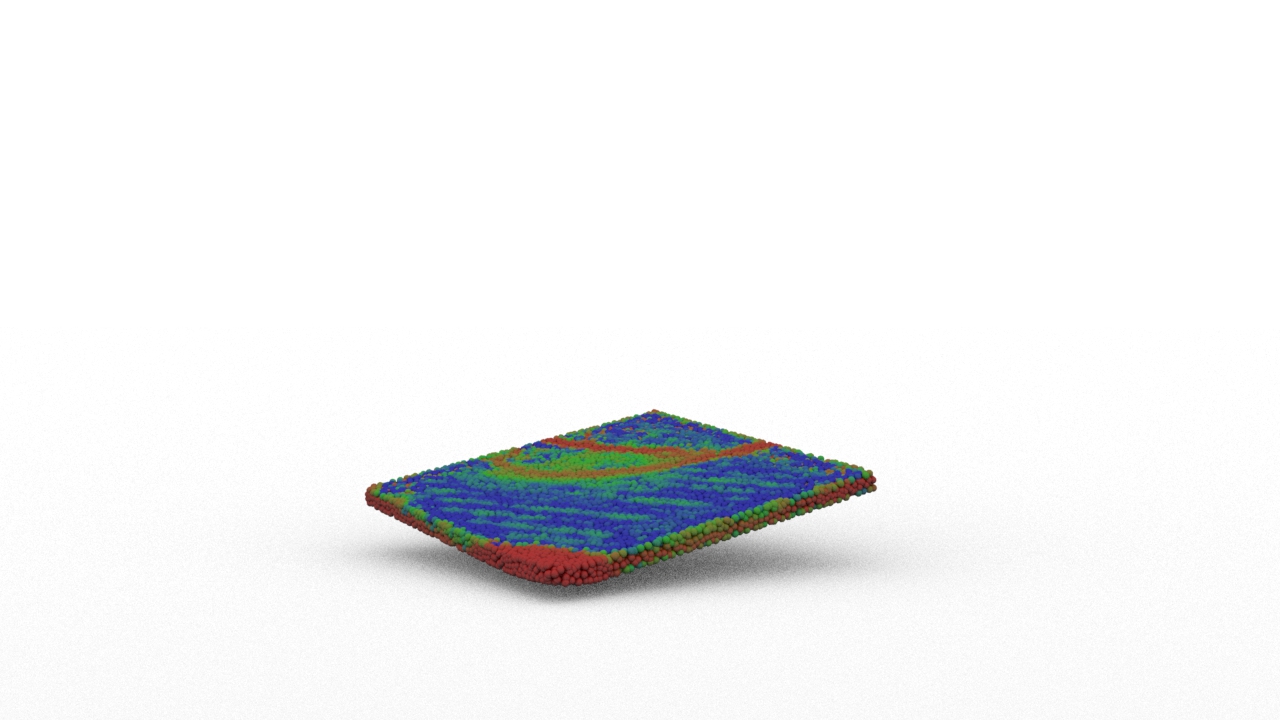}
    \caption{This figure provides a visualization of the damage profile with MPM. It is evident that, with the correct material strength parameters applied, the chocolate material exhibits damage effects under high-velocity impact conditions. However, this setting does not precipitate fracture, necessitating the introduction of further rules and controls to direct the fragmentation process. This requirement highlights a limitation inherent to such continuum-based methods, where additional rules are needed to simulate realistic behavior.}
    \label{fig:chocolate_mpm_viz}
\end{figure}

\paragraph{Chocolate Falling}
The chocolate falling experiment was primarily designed to demonstrate the distinctions in simulation outcomes among the three methods under high-velocity impact conditions. In the experiment, the chocolate was consistently dropped onto a floor surface at a velocity of $5 m/s$. The fracture results of the three methods are presented in \Cref{fig:chocolate_compare}. The baseline FEM exhibits severe "shattering" artifacts. Extensive parameter tuning yields either numerous tiny fragments or a few large pieces, failing to produce the desired intermediate fragmentation pattern with diverse fragment sizes. Figure \ref{fig:chocolate_mpm_viz} presents a detailed visualization of the damage phase field in MPM. While the MPM method captures the continuum damage of materials, it has failed to demonstrate a plausible fragmentation process. This limitation stems from the constraints of the phase field model, which necessitates the introduction of additional artificial rules and parameters for artistic control over cracking and fragmentation within this scenario \cite{wolper2019cd,fan2022simulating}. While it is conceivable that the introduction of further supplementary rules might enhance the simulation outcomes for this particular case, it is clear that such modifications are extrinsic to the material's fundamental properties. In contrast, our method necessitates only the specification of material parameters to simulate the fracture phenomenon, thus achieving a realistic and authentic fragmentation effect.

\begin{table*}[ht]
    \centering
    \caption{The simulation statistics for our experiments are listed below, with the slow-motion of 1x corresponding to one frame per 1/60th of a second, and 0.125x corresponding to one frame per 1/480th of a second. The time step and time consumption for the proposed approach are compared with explicit and implicit methods at different element numbers and stiffness settings.}
    \begin{tabular}{ccccccccc}
    \thickhline
    Scene & Method & \# points & $\Delta t$(s) & $E$ & $\nu$ & $\tau$ & time per frame (s) & fracture frame \\ 
    \thickhline
    \multirow{3}{*}{Three Point Bending (Low)} & FEM & 578 & 4.2E-5 & 1E7 & 0.3 & 300 & 6.2 & 6 \\ \cline{2-9}
                                               & MPM & 500 & 4.2E-5 & 1E7 & 0.3 & 3E4 & 0.6 & 26 \\ \cline{2-9}
                                               & Ours & 532 & 2.0E-5 & 1E7 & 0.3 & 3E4 & 0.9 & 23\\ \hline
    \multirow{3}{*}{Three Point Bending (Middle)} & FEM & 1.7K & 2.8E-5 & 1E7 & 0.3 & 300 & 57.6 & 13 \\ \cline{2-9}
                                                  & MPM & 2K & 2.8E-5 & 1E7 & 0.3 & 3E4 & 5.2 & 23\\ \cline{2-9}
                                                  & Ours & 1.8K & 9.1E-6 & 1E7 & 0.3 & 3E4 & 7.3 & 25 \\ \hline
    \multirow{3}{*}{Three Point Bending (High)} & FEM & 16.4K & 2.0E-5 & 1E7 & 0.3 & 300 & 1240 & 29 \\ \cline{2-9}
                                                & MPM & 20K & 2.1E-5 & 1E7 & 0.3 & 3E4 & 48.2 & 11 \\ \cline{2-9}
                                                & Ours & 18K & 5.8E-5 & 1E7 & 0.3 & 3E4 & 137.1 & 25 \\ \hline
    \multirow{3}{*}{Chocolate Falling (0.125x slow motion)} & FEM & 13.6K & 5.2E-6 & 1E8 & 0.3 & 100 & 650.1 & - \\ \cline{2-9}
                                       & MPM & 20K & 4.2E-6 & 1E8 & 0.3 & 5E4 & 24.5 & - \\ \cline{2-9}
                                       & Ours & 22K & 4.4E-5 & 1E8 & 0.3 & 5E4 & 106.8 & - \\ \hline
    \thickhline
    \end{tabular}
    \label{tab:comparison_params}
\end{table*}

\Cref{tab:comparison_params} details the computational time costs and settings for both experiments. The data reveals that BDEM achieves more consistent fracture frames across resolutions compared to FEM and MPM, indicating fracture occurs at similar deformation states and demonstrating superior scale consistency. 

Per-frame computational costs indicate FEM is the slowest method, requiring residual propagation for accurate crack patterns, resulting in more iterations per step. While MPM has lower per-frame computational costs than BDEM, its inferior scale consistency necessitates extensive parameter tuning of material strength and grid resolution to achieve satisfactory fracture results. In contrast, BDEM simplifies this process significantly: suitable material strength parameters can be rapidly determined at low resolution and then applied directly to medium and high-resolution simulations with consistent results.

Analysis of fracture thresholds reveals an intriguing discrepancy. Despite all three methods employing thresholds with identical physical dimensions, the numerical value for the FEM threshold diverges significantly from the other two approaches. Notably, this smaller threshold value in FEM aligns with parameters in the original work by \citet{o1999graphical}, while BDEM and MPM thresholds more closely approximate real-world material strengths.

\subsection{Comparison with Explicit Method}
\paragraph{Beam Stretch \& Twist}
In our experiments, we compared the simulation results of our implicit method with the explicit method proposed by \cite{lu2021simulating}. Specifically, we evaluated their performance in beam stretching and twisting scenarios, and found that both methods produced identical results under the same settings.

\begin{figure}[htbp]
    \centering
    \includegraphics[width=0.49\linewidth]{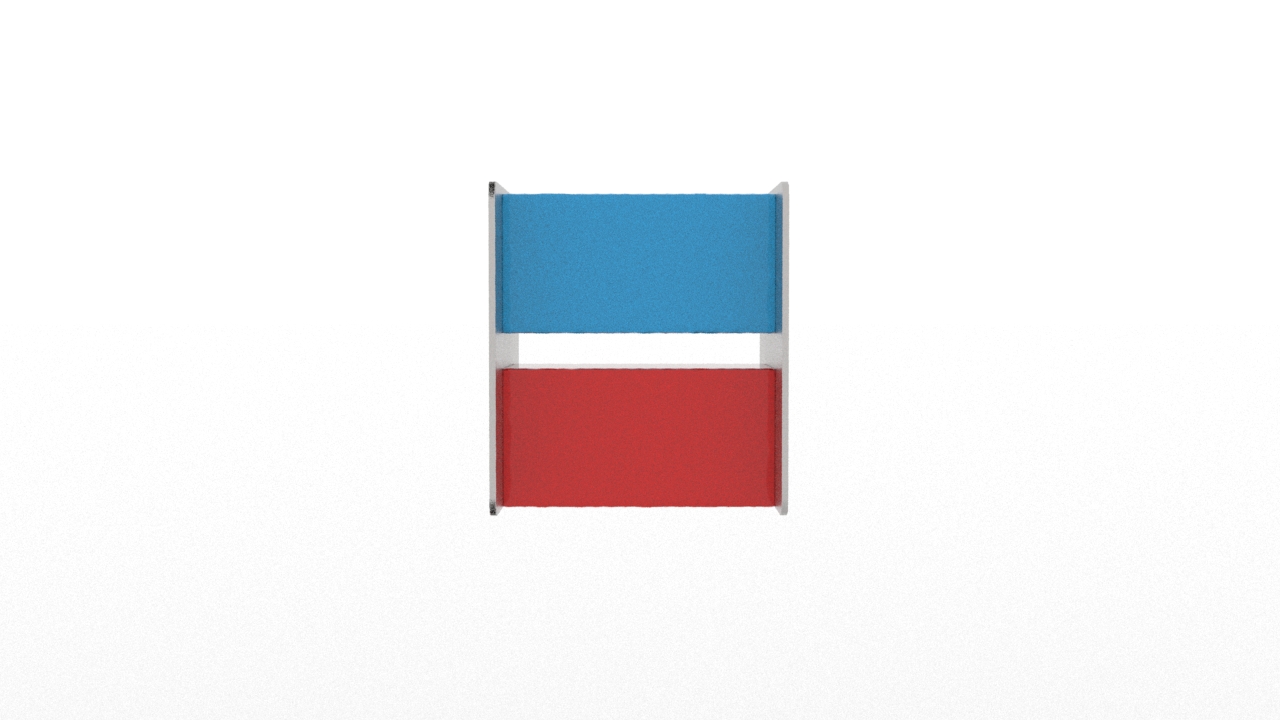}
    \includegraphics[width=0.49\linewidth]{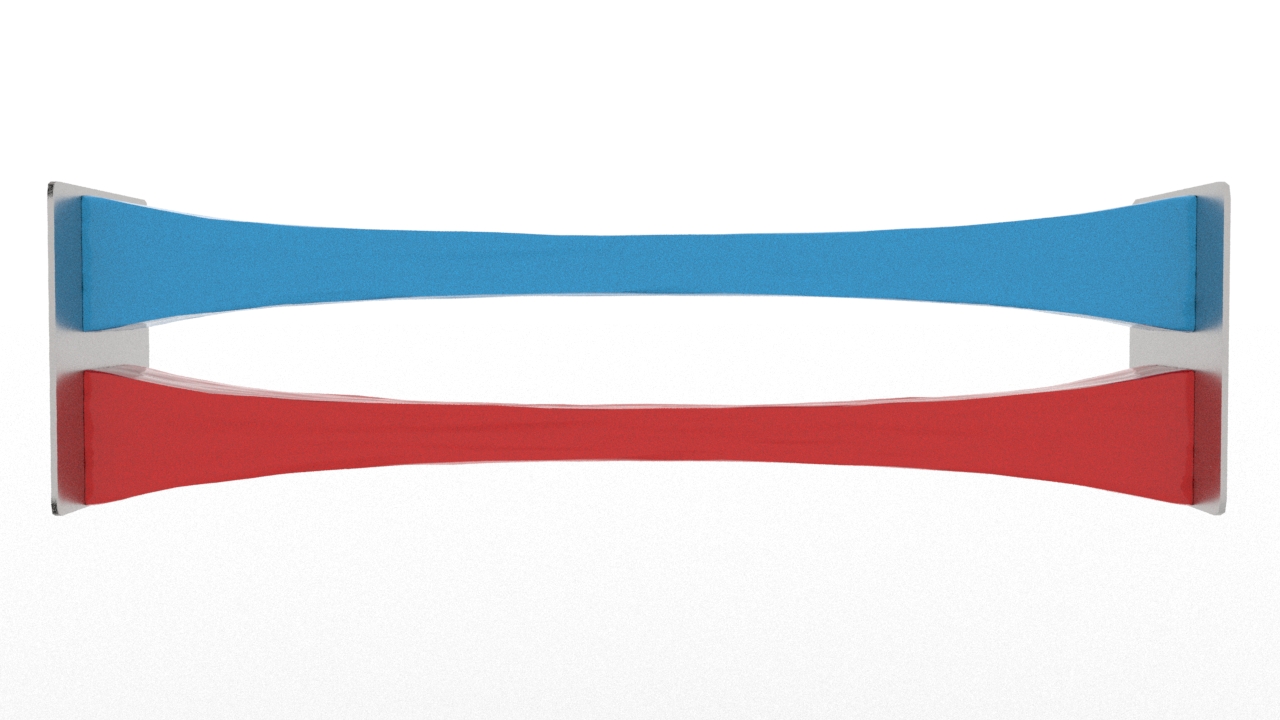}
    \includegraphics[width=0.49\linewidth]{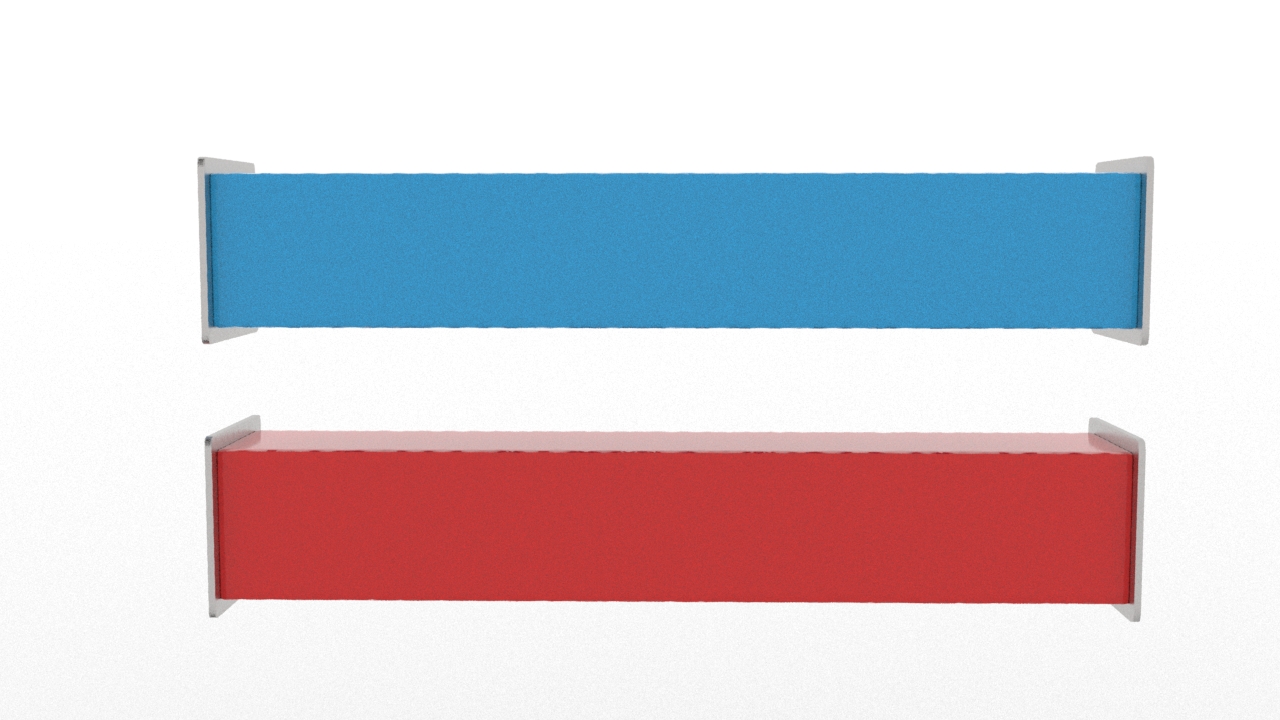}
    \includegraphics[width=0.49\linewidth]{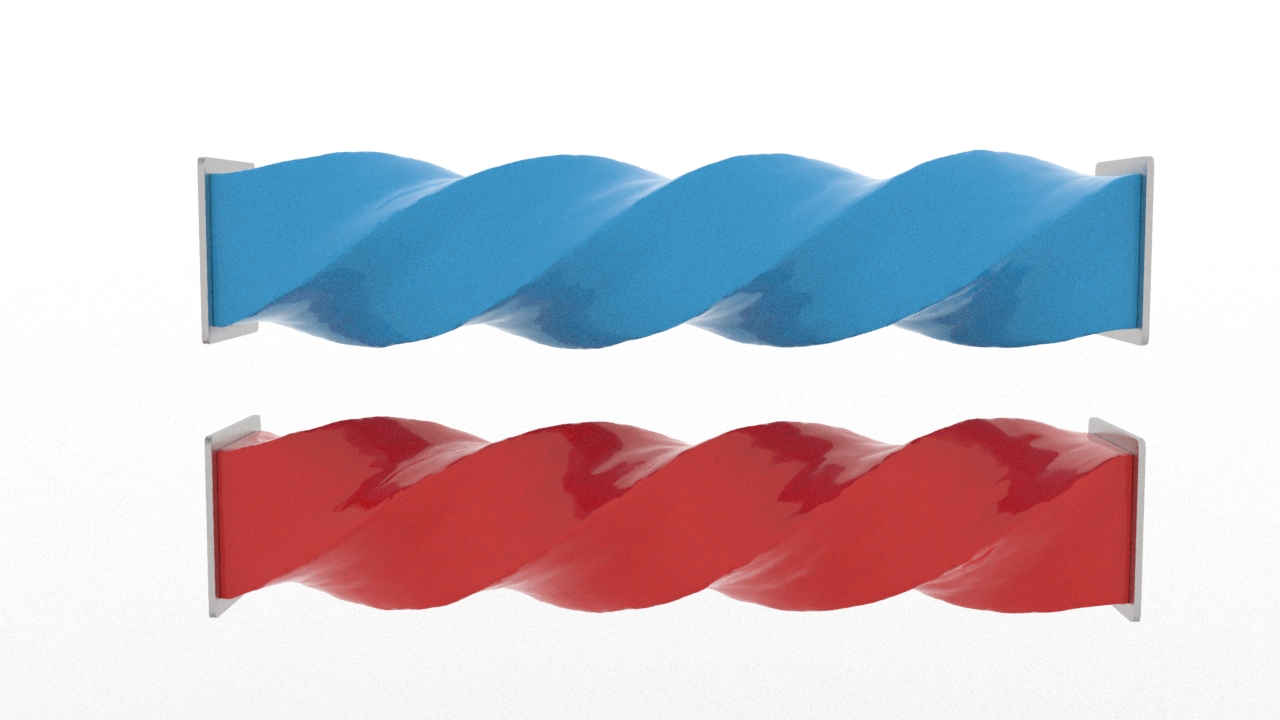}
    \caption{Comparison of explicit (blue) and implicit (red) methods in simulating a stretched and twisted beam. Both approaches yield similar results.}
    \label{fig:beam_stretch_twist}
\end{figure}

\begin{figure}[htbp]
    \centering
    \includegraphics[width=0.29\linewidth]{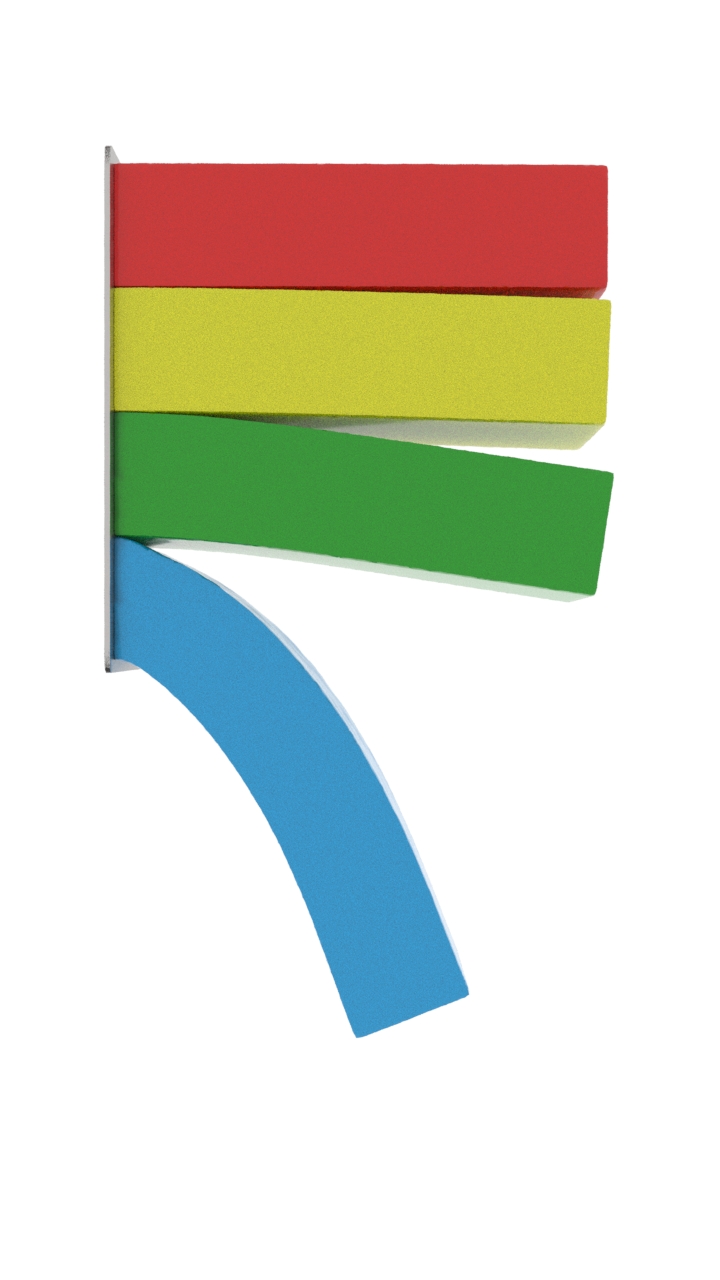}
    \includegraphics[width=0.7\linewidth]{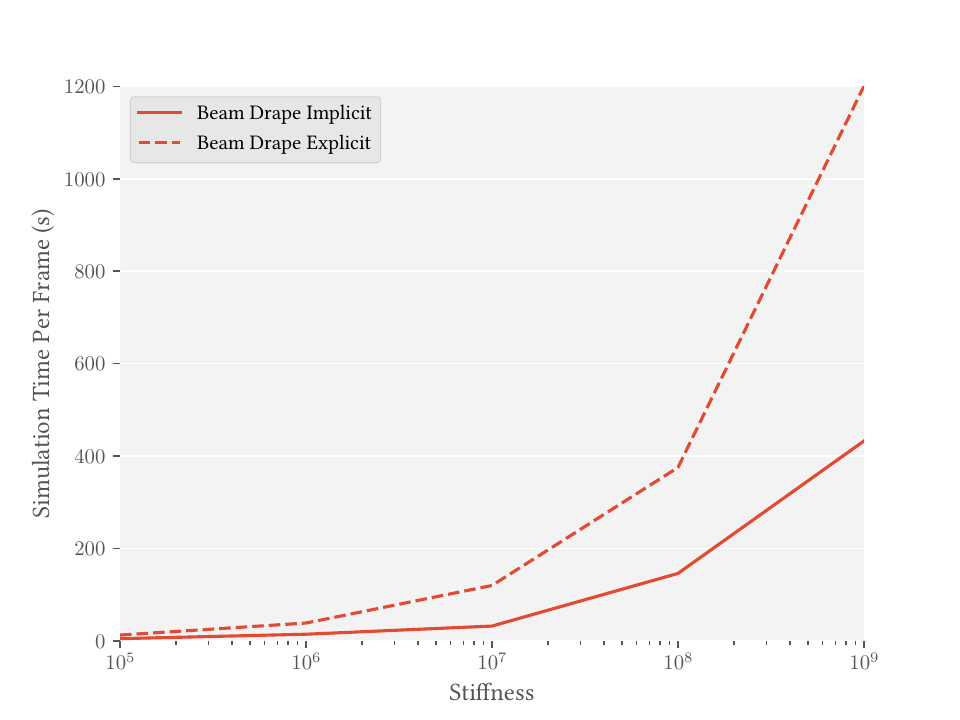}
    \caption{This experiment compares explicit and implicit methods for simulating beam drape. The left images illustrate static drape scenarios with varying material stiffness: red ($10^9$), yellow ($10^8$), green ($10^7$), and blue ($10^6$). The curves in the right image plot the relationship between computational cost and material stiffness, demonstrating that implicit methods produce faster results for a range of stiffness values.} 
    \label{fig:beam_drape_stiffness}
\end{figure}

\paragraph{Beam Drape}
This experiment investigates and compares the simulation time of the implicit and explicit simulation methods using the beam drape scenario with different stiffness levels. Two curves are plotted in Figure \ref{fig:beam_drape_stiffness} to examine the relationship between simulation time and stiffness, as well as simulation time and scale. Our results show that the implicit method outperforms the explicit method by $3$ to $6$ times in terms of run time.

\begin{figure}[htbp]
    \centering
    \includegraphics[width=\linewidth]{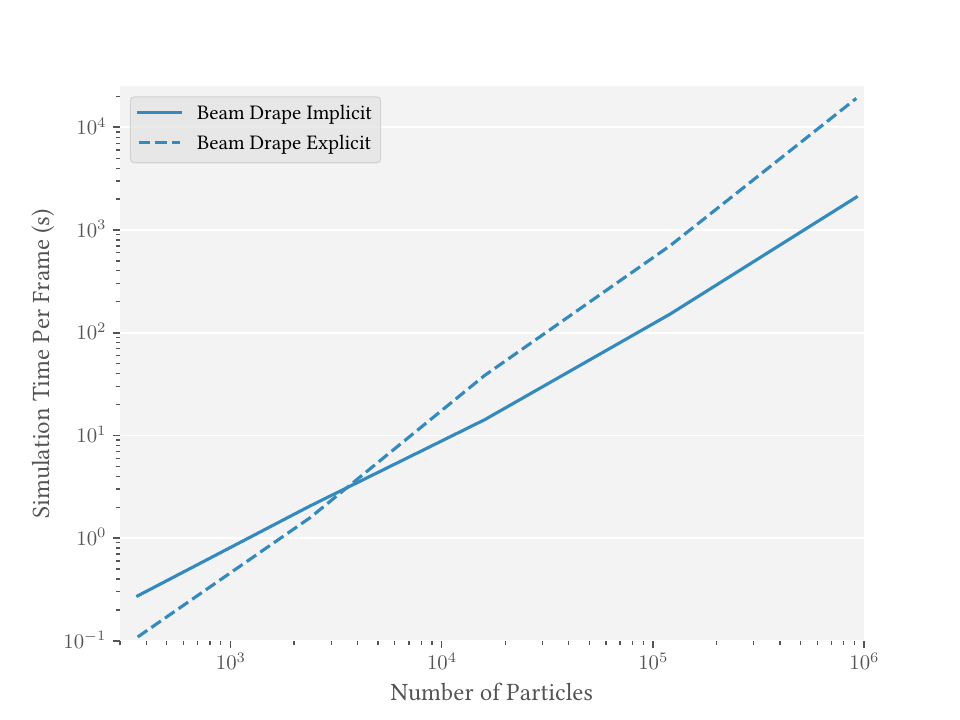}
    \caption{This experiment compares explicit and implicit methods for simulating beam drape. The curves plot the relationship between the computational cost and simulated discrete elements, showing implicit methods significantly outperform explicit methods for finer simulation scales.}
    \label{fig:beam_drape_scale}
\end{figure}

\paragraph{Chocolate}
In this experiment, the deformation of a chocolate drop with high stiffness is simulated to evaluate the computational efficiency between implicit and explicit simulation methods in rich contact scenarios. The outcomes indicate that the implicit method outperforms the explicit method by achieving a 2.5x speedup. The visualization of the simulation can be inspected in Figure \ref{fig:chocolate}.

\paragraph{Plate}
This experiment compares the computational efficiency of implicit and explicit methods for simulating the fracture of a plate with high stiffness upon impact with the ground. Our method has demonstrated a 2.4x increase in computational speed relative to the explicit method under these conditions. The simulation results are illustrated in Figure \ref{fig:plate}.

\paragraph{Cloth}
Here, we examine the computational efficiency of implicit versus explicit methods in the context of yarn-level cloth simulation, characterized by high tensile and low bending stiffness of the yarns. Our method provides a 3.3x computational speed advantage over the explicit approach under these circumstances. The dynamics of the cloth tearing, as simulated, are depicted in Figure \ref{fig:cloth}.

\section{Conclusion}
\label{sec:Conclusion}

We present an optimization-based implicit integrator for BDEM system, and introduce a manifold optimization scheme to transform the nonlinear dynamic simulation into an unconstrained optimization problem on a spherical manifold, which significantly improves the solution stability and efficiency. In addition, a nullspace operator is proposed, which simplifies the optimization procedure and reduces the number of unknowns. Additionally, our new implicit BDEM approach uses the semi-positive definite projection method with preconditioned conjugate gradient and line search method on the manifold, which achieves a stable integrator. In our experiments, we have showcased that our method surpasses the explicit BDEM in terms of computational speed. Furthermore, when compared to FEM and MPM for fragmentation, our approach exhibits superior scale consistency, accurately simulating fracture effects at smaller scales. The simulation of fragmentation is more realistic and natural with our method, which necessitates only the provision of material parameters to get the authentic fracture process.

While our approach offers notable advancements, it is not without its limitations. Specifically, to mitigate complexities at the contribution points, we have utilized a rudimentary collision model for collision management. This simplification may lead to a diminished fidelity in the representation of collision effects. In future work, we plan to integrate more sophisticated collision models into our optimization framework to enhance the accuracy and realism of our simulations. Furthermore, the BDEM system we have developed is highly amenable to parallelization. Although our current exposition does not delve into the details of GPU acceleration for the sake of a clear computational comparison, we intend to explore a more detailed implementation of parallel acceleration in our methods moving forward. Additionally, while our approach successfully captures fine-grained collision details, the reconstruction and rendering of fracture surfaces also play a crucial role in the final visual outcomes. We aim to explore fracture surface reconstruction methods more tailored to the nuances of BDEM in our ongoing research.

% %%
% %% The acknowledgments section is defined using the "acks" environment
% %% (and NOT an unnumbered section). This ensures the proper
% %% identification of the section in the article metadata, and the
% %% consistent spelling of the heading.
% \begin{acks}
% \end{acks}

\bibliographystyle{ACM-Reference-Format}
\bibliography{sample-base}

\appendix

\section{Quaternion Operations}
\label{sec:Quaternion Operations}

Quaternions are generally represented in the following form,
\begin{equation}
    q = u \mathbf{i} + v \mathbf{j} + w \mathbf{k} + s = \{ u, v, w, s \} = \begin{pmatrix}
        u \\ v \\ w \\ s
    \end{pmatrix} ,
\end{equation}
where $u$, $v$, $w$ and $s$ are real numbers, and 1, $\mathbf{i}$, $\mathbf{j}$, and $\mathbf{k}$ are the basis vectors or basis elements. A unit quaternion can denote a rotation in three-dimensional space. Compared to other forms of rotation representations such as Euler angles, axis-angle and rotation matrices, quaternions have advantages in computational simplicity and efficiency. Hence, quaternions are widely used in computer graphics to represent rotation, including BDEM \cite{lu2021simulating} where they are used to describe the rotational status of discrete elements. As quaternions play a key role in formulating the proposed implicit BDEM, we briefly recap the key information related to quaternion operations in this section. 

\paragraph{Quaternion Multiplication}
The multiplication of quaternions represents the composition of rotations: the left multiplication represents the order of rotations, and the right multiplication represents the subsequent rotation of the coordinate system. Rotations can also be represented in the matrix form with quaternions: 
\begin{equation}
    q_i q_j = Q_l(q_i) q_j = Q_r(q_j) q_i ,
\end{equation}
where $Q_l(q)$ is called the left multiplication matrix and $Q_r(q)$ the right multiplication matrix. Let vector $t = \{u, v, w\}$ denote the imaginary part of quaternion $\{ u, v, w, s \}$, the left and right multiplication matrices can be expressed as follows: 
\begin{equation}
    Q_l(q) = 
    \begin{pmatrix}
        s I_3 + t^{\times} & t \\
        -t^T & s
    \end{pmatrix}
\end{equation}
\begin{equation}
    Q_r(q) = 
    \begin{pmatrix}
        s I_3 - t^{\times} & t \\
        -t^T & s
    \end{pmatrix} ,
\end{equation}
where the skew symmetric matrix $t^{\times}$ is the cross matrix of vector $t$
\begin{equation}
    t^{\times} = 
    \begin{pmatrix}
         0 & -w &  v \\
         w &  0 & -u \\
        -v &  u &  0
    \end{pmatrix} .
\end{equation}

\paragraph{Direction Rotation Operator}
Let $d_0$ denote a direction in the 3D space, the new direction $d$ after a rotation can be expressed as: 
\begin{equation}
    \begin{pmatrix}
        d \\ 0
    \end{pmatrix}
    = q \begin{pmatrix}
        d_0 \\ 0
    \end{pmatrix} \overline{q}
    \label{eq:rotation_1} ,
\end{equation}
where quaternion $q$ denotes the rotation and $\overline{q} = \{-u, -v, -w, s\}$ is its conjugate. The above rotation operation can be rewritten as:  
\begin{equation}
    d = q \odot d_0,
\end{equation} 
where $\odot$ is the direction rotation operator. 

\paragraph{Nullspace Operator}
\label{appendix:nullspace_operator}
The composition of a quaternion its and conjugate is the unit quaternion, i.e. $\overline{q} q = \{0, 0, 0, 1\}$. In the matrix form, this becomes 
\begin{equation}
    \overline{q} q = Q_l(\overline{q}) q = 
    \begin{pmatrix}
        s I_3 - t^{\times} & -t \\
        t^T & s
    \end{pmatrix}
    \begin{pmatrix}
        t \\
        s
    \end{pmatrix} = 
    \begin{pmatrix} 0 \\ 0 \\ 0 \\ 1 \end{pmatrix} .
\end{equation}
We can define the upper matrix of $Q_l(\overline{q})$ as a nullspace operator: 
\begin{equation}
    G_l(q) =
    \begin{pmatrix}
        sI_3 - t^{\times}& -t
    \end{pmatrix}, 
    G_r(q) =
    \begin{pmatrix}
        sI_3 + t^{\times}& -t
    \end{pmatrix} .
\end{equation}
The nullspace operator can be used to reduce the computational cost of quaternion systems and the detailed strategy is provided in \Cref{subsec:second_order_chap}. With the nullspace operator, the following relations hold: 
\begin{equation}
    G_l(q) q = 0, G_r(q) q = 0,
\end{equation}
and
\begin{equation}
\label{eqn:nullspace_property}
\begin{split}
    G_l(q)^TG_l(q) &= |q|^2I_4-q q^T ,\\
    G_l(q)G_l(q)^T &= |q|^2I_3 .
\end{split}
\end{equation}
The multiplication $G_l(q_i) q_j$ removes the component of $q_i$ in $q_j$ and projects the vector to the direction of angle axis of quaternion $\overline{q_i} q_j$. The quaternion matrix can also be related to the nullspace operator matrix as
\begin{equation}
    Q_l(q) = \begin{pmatrix}
        G_l(q)^T & q
    \end{pmatrix},
    Q_r(q) = \begin{pmatrix}
        G_r(q)^T & q
    \end{pmatrix} .
\end{equation} 

\paragraph{Derivation and Angular Velocity}
Let $w_4$ denote the angular velocity in four dimensions: 
\begin{equation}
    w_4=\begin{pmatrix}
        w \\
        0\\
    \end{pmatrix} .
\end{equation}
The derivative of a quaternion and the extended angular velocity $w_4$ satisfy the following relations \cite{betsch2009rigid}:
\begin{equation}
    \dot{q} = \frac{1}{2} w_4 q ,
\end{equation}
\begin{equation}
    \label{eqn:extend_angular_velocity}
    w_4 = 2 \dot{q} \overline{q} .
\end{equation}

\paragraph{Quaternion Constraint}
A rotation can be represented by a unit quaternion. To ensure the quaternion remains as a unit during calculations, the unit constraint must be applied: 
\begin{equation}
    \label{eqn:quaternion_constraint}
    C(q)=\frac{1}{2}(q^Tq-1)=0 .
\end{equation}

\section{Hamiltonian System with Backward Euler Formulation}
\label{sec:Hamilton_System}
We present a derivation of the kinetic equation for rotational systems using a conventional approach: first establishing the continuous formulation, then applying a time discretization method to obtain the integrator form.
As described by \cite{betsch2009rigid}, the continuous form of the Hamiltonian system equation is given by:
\begin{equation}
\begin{split}
\dot{q} &= \frac{1}{4}(\mathbb{J}^{-1}_0(q\cdot p)q+G(q)^T J^{-1}G(q)p),\\
\dot{p} &= -\frac{1}{4}(\mathbb{J}^{-1}_0(q\cdot p)p+G(p)^T J^{-1}G(p)q)-\nabla V(q),
\end{split}
\end{equation}
where $q$ represents the rotational state, $p$ denotes the extended momentum, and $\mathbb{J}_0=\text{tr}(J)$. Given the quaternion unit length constraint with property $q\cdot p = 0$ \cite{betsch2009rigid}, we can simplify the equation to:
\begin{equation}
\begin{split}
\dot{q} &= \frac{1}{4}(G(q)^T J^{-1}G(q)p)\\
\dot{p} &= -\frac{1}{4}(G(p)^T J^{-1}G(p)q)-\nabla V(q)
\end{split}
\end{equation}
For a spherical element, $J = \frac{2}{5}mr^2 I_3$. Let $\mathbb{I} = \frac{2}{5}mr^2$. Applying the result from \Cref{eqn:nullspace_property}, we further simplify the equation:
\begin{equation}
\begin{split}
\dot{q} &= \frac{1}{4\mathbb{I}}((I_4 - qq^T) p),\\
\dot{p} &= -\frac{1}{4\mathbb{I}}(|p|^2 I_4 - pp^T) q-\nabla V(q).
\end{split}
\end{equation}
Maintaining $q\cdot p=0$, we obtain a more concise form:
\begin{equation}
\begin{split}
\dot{q} &= \frac{1}{4\mathbb{I}}p\\
\dot{p} &= -\frac{1}{4\mathbb{I}}|p|^2 q-\nabla V(q)
\end{split}
\end{equation}
Integrating the first equation into the second and simplifying, we derive an equation solely in terms of $q$:
\begin{equation}
4\mathbb{I}(\ddot{q}+|\dot{q}|^2q) + \nabla V(q) = 0
\end{equation}
Note the new term $|\dot{q}|^2q$, which distinguishes this formulation from non-rotational cases. To align with our manifold optimization approach, we left-multiply by $(I_4 - qq^T)$. Since $(I_4 - qq^T) q = 0$, the residual on the manifold becomes:
\begin{equation}
(I_4 - qq^T) (4\mathbb{I}\ddot{q} + \nabla V(q)) = 0
\end{equation}
Observe that $4\mathbb{I}\ddot{q} + \nabla V(q)$ and $4\mathbb{I}(\ddot{q}+|\dot{q}|^2q) + \nabla V(q)$ yield identical residuals on the manifold. Consequently, we can employ the simpler form, and discretization using backward Euler results in the same expression as in \Cref{eqn:implicit_euler_integrator}.

\section{Discrete Lagrangian}
\label{sec:discrete_lagrangian}
Hamilton's principle is a fundamental relation in physics to describe the motion of a dynamic system. By taking the variation of the Lagrangian functional $S = \int_{t_0}^{t_1} L(x(t),\dot x(t)) dt$, the equations of motion can be obtained as the stationary point of the action functional. For our system, the Lagrangian function given by the sum of the  kinetic energy $T$ and the potential energy $V$: 
\begin{equation}
L(x,\dot{x})=T(x,\dot{x})-V(x,\dot{x})=\frac{1}{2}\dot{x}^TM\dot{x}-V(x).
\end{equation}
The true evolution of the physical system satisfies  $\frac{\partial{S}}{\partial{x(t)}} = 0$. For quaternion-based Hamiltonian system, we can articulate the equations within which the extended kinetic energy form is derivable from \cite{betsch2009rigid}. 
\begin{equation}
    T(q,\dot{q}) = \dot{q}^TM_q(q)\dot{q}.
\end{equation} 
Here, for one element, $M_q(q) = 4G_l(q)JG_l(q)^T+2\text{tr}(J)qq^T$, $J$ is the moment of inertia matrix and $G_l(q)$ is the nullspace operator in appendix \ref{appendix:nullspace_operator}. Given that the mass matrix $M_q(q)$ varies with $q$, direct differentiation of the equations yields a form akin to the extended momentum in rigid body motion, which is complicated for the derivation of an incremental potential form. In this context, we employ the Discrete Lagrangian formulation to re-derive the equations and perform simplifications based on some special properties of the BDEM.

In a small time interval, the action functional can be approximated by a discrete Lagrangian: 
\begin{equation}
\label{eqn:discrete_lagrangian}
S_d=\sum_{i=0}^{n-1}L_d(t_i,t_{i+1},x_i,x_{i+1}) \approx\int_{t_0}^{t_{n}}L(t, x(t))dt.
\end{equation}
At the stationary point of the above discrete action functional, each term of the sum must be set to zero: 
\begin{equation}
\label{eqn:stationary}
\frac{\partial{S_d}}{\partial{x_i}}=0=\frac{\partial L_d(t_{i-1},t_i,x_{i-1},x_i)}{\partial x_i}+\frac{\partial L_d(t_i,t_{i+1},x_i,x_{i+1})}{\partial x_i}.
\end{equation}

Here, with mid-point approximation, 
\begin{equation}
\begin{split}
T_d(x_i, v_i) &= \Delta t T(\frac{x_{i+1}+x_i}{2}, \frac{x_{i+1}-x_i}{\Delta t})\\
V_d(x_i) &= \Delta t V(\frac{x_{i+1}+x_i}{2}).
\end{split}
\end{equation}

While this formulation is general for any element shape, our BDEM system employs spherical discrete elements exclusively. This choice allows us to leverage the symmetry properties of spheres, significantly simplifying the computational process. It's important to note that these optimizations are specific to spherical elements and do not apply to other geometries.

For a sphere with mass $m$ and radius $r$, the mass matrix $M_p = m I_3$, whilst its moment of inertia is $J = \frac{2}{5}mr^2 I_3$, here $I_3$ stands for the $3\times3$ identity matrix. With vector $p$ denotes its position and quaternion $q$ its rotation, the discrete energy term for position, can be easily computed with linear velocity approximation:
\begin{equation}
    T_d^p = \Delta t\frac{1}{2}(\frac{p_{i+1}-p_i}{\Delta t})^T m I_3 (\frac{p_{i+1}-p_i}{\Delta t}).
\end{equation}
For rotation, we approximate the mass matrix and velocity by a mid-point approximation:
\begin{equation}
    T_d^q = \Delta t\frac{1}{2}(\frac{q_{i+1}-q_i}{\Delta t})^T M_q(\frac{q_{i+1}+q_i}{2})(\frac{q_{i+1}-q_i}{\Delta t}).
\end{equation}
With the definition of extended matrix and moment of inertia for sphere, we can get:
\begin{equation}
    \begin{split}
        M_q(q) &= 4G(q)JG(q)^T+2\text{tr}(J)qq^T \\
               &= \frac{8}{5}mr^2(I_4-qq^T)+\frac{12}{5}mr^2 qq^T\\
               &= \frac{8}{5}mr^2(I_4+\frac{1}{2}qq^T),
    \end{split}
\end{equation}
here $I_4$ stands for the $4\times 4$ identity matrix. With unit length property of quaternion during our simulation, we have $(q_{i+1}+q_i)^T(q_{i+1}-q_i)=q_{i+1}^Tq_{i+1}-q_i^Tq_i=0$. With the mid-point mass matrix as in \cite{betsch2009rigid}, the discrete kinetic energy is:
\begin{equation}
    \begin{split}
        T_d^q &= \frac{1}{2\Delta t}(q_{i+1}-q_i)^T M_q(\frac{q_{i+1}+q_i}{2})(q_{i+1}-q_i)\\
        & = \frac{1}{2\Delta t}(q_{i+1}-q_i)^T \frac{8}{5}mr^2(I_4 + \frac{1}{2}\frac{q_{i+1}+q_i}{2}(\frac{q_{i+1}+q_i}{2})^T)(q_{i+1}-q_i) \\
        & = \frac{1}{2\Delta t}(q_{i+1}-q_i)^T \frac{8}{5}mr^2 I_4(q_{i+1}-q_i).
    \end{split}
\end{equation}
Now the mass matrix for quaternion becomes constant here. The extended mass matrix $M_s$ is
\begin{equation}
    M_s=\begin{pmatrix}
        m I_3 &\\
        & \frac{8}{5}mr^2 I_4
    \end{pmatrix}.
\end{equation}
Concatenating positions and rotations for all elements as system state $x$, the discrete kinetic energy of the spherical element can be expressed as: 
\begin{equation}
    T_d = \frac{1}{2}\Delta t(\frac{x_{i+1}-x_i}{\Delta t})^TM_s(\frac{x_{i+1}-x_i}{\Delta t}),
\end{equation}
The discrete Lagrangian can be computed by
\begin{equation}
L_d(t_i,t_{i+1},x_i,x_{i+1})=\Delta t \frac{1}{2}(\frac{x_{i+1}-x_i}{\Delta t})^TM_s(\frac{x_{i+1}-x_i}{\Delta t})-\Delta t V(\frac{x_{i+1}+x_i}{2}),
\end{equation}
the terms in Eq. (\ref{eqn:stationary}) can be expressed as:
\begin{equation}
    \frac{\partial L_d(t_{i-1},t_i,x_{i-1},x_i)}{\partial x_i} = \frac{1}{\Delta t}M_s(x_i-x_{i-1})
    -\frac{1}{2}\Delta t\nabla V(\frac{x_{x-1}+x_i}{2}),
\end{equation}
\begin{equation}
    \frac{\partial L_d(t_i,t_{i+1},x_i,x_{i+1})}{\partial x_i} = \frac{1}{\Delta t}M_s(x_i-x_{i+1})
    -\frac{1}{2}\Delta t\nabla V(\frac{x_i+x_{i+1}}{2}).
\end{equation}
Let $\hat x = 2x_i-x_{i-1}$, the mid-point form of our variational integrator can be obtained as: 
\begin{equation}
    \frac{1}{\Delta t^2}M_s(x_{i+1}-\hat x)+\frac{1}{2}\nabla V(\frac{x_{i+1}+x_i}{2})+\frac{1}{2}\nabla V(\frac{x_{i-1}+x_i}{2})=0.
    \label{eqn:variational_integrator_original}
\end{equation}
Now we have developed a variational integrator for BDEM system simulations. The following part outlines the derivation of our optimization formulation, which forms the foundation of our subsequent optimization process.
We introduce a new variable $\hat{z} = \hat x - \frac{1}{2}M_s^{-1}\Delta t^2\nabla V(\frac{x_{i-1}+x_i}{2})$, calculable from known previous states. This allows us to reformulate the equation as:
\begin{equation}
\frac{1}{\Delta t^2}M_s(x_{i+1}-\hat{z})+\frac{1}{2}\nabla V(\frac{x_{i+1}+x_i}{2})=0.
\label{eqn:variational_integrator}
\end{equation}
Observing this equation, we can recast it as an optimization problem with the objective function:
\begin{equation}
\label{eqn:variational_optimization_form}
\Phi(\bm x_{i+1}) = \frac{1}{2\Delta t^2}(x_{i+1}-\hat{z})^T M_s (x_{i+1}-\hat{z})+V(\frac{x_{i+1}+x_i}{2}).
\end{equation}

\end{document}